\documentclass[3p,sort&compress]{elsarticle}

\usepackage{amssymb}    
\usepackage{amsmath}  
\usepackage{subfigure}  
\usepackage{epstopdf}   
\usepackage[separate-uncertainty=true]{siunitx}   
\usepackage{color}      
\usepackage{soul}       
\usepackage{lmodern}    
\usepackage{hyperref}   
\usepackage{listings}   
\usepackage{amsmath}
\usepackage{amsmath}
\usepackage{amssymb}
\usepackage{graphicx}
\usepackage{epstopdf}
\usepackage{booktabs}

\usepackage[utf8]{inputenc}
\usepackage[T1]{fontenc}
\usepackage{amssymb}    
\usepackage{amsmath}  
\usepackage{subfigure}  
\usepackage{epstopdf}

\usepackage[most]{tcolorbox}
\newtcolorbox{highlighted}{colback=yellow,coltext=red,breakable}
\usepackage{academicons}
\usepackage{xcolor}

\newcommand{\orcid}[1]{\href{https://orcid.org/#1}{\textcolor[HTML]{A6CE39}{\aiOrcid}}}

\begin{document}

\title{Evolution of Quantum Nonequilibrium for Coupled Harmonic Oscillators}

\author{Francisco Bento Lustosa$^{1}$, Nelson Pinto-Neto$^{1}$, Antony Valentini$^{2}$}

\address{$^{1}$Centro Brasileiro de Pesquisas Físicas,\\
Rua Dr. Xavier Sigaud 150, Urca - CEP: 22290-180 - Rio de Janeiro-RJ,  Brazil\\
$^{2}$Augustus College, 14 Augustus Road, London SW19 6LN, UK.\\
Department of Physics and Astronomy,\\
Clemson University, Kinard Laboratory,\\
Clemson, SC 29634-0978, USA.}

\begin{abstract}
In the context of de Broglie-Bohm pilot-wave theory, violations of the Born rule are allowed and can be considered as describing nonequilibrium distributions. We study the effects of interactions on quantum relaxation towards equilibrium  for a system of one-dimensional coupled harmonic oscillators. We show by numerical simulations that interactions can delay or even prevent complete relaxation for some initial states. We also discuss how this effect might be relevant for cosmological scenarios and how nonequilibrium could be detected in some models. 
\end{abstract}

\maketitle

\section{Introduction} \label{sec:introduction}

The de Broglie-Bohm pilot-wave theory provides a description of quantum systems in terms of trajectories (or configurations $q(t)$) that evolve continuously through time guided by a wave function given by the usual solution of the Schrödinger equation \cite{bib:deBroglie1928, bib:Bacciagaluppi2006, bib:Bohm1951a, bib:Bohm1951b, bib:Holland1995}. In that context, quantum probability distributions can deviate from the usual Born rule and, through a process of `quantum relaxation', those distributions can relax to the state of equilibrium $\rho(q,t) = |\psi(q,t)|^2$  \cite{bib:Valentini1990, bib:Valentini1992, bib:Valentini2001, bib:Valentini2019, bib:Valentini2005, bib:Efthymiopoulos2006, bib:Towler2012, bib:Colin2011, bib:Abraham2014, bib:Efthymiopoulos2017, bib:Lustosa2020}. This process is analogous to thermal relaxation and can be described in terms of a `subquantum \textit{H}-theorem'. Considering that all experimentally observed quantum systems respect the Born rule and, therefore, are in quantum equilibrium it is natural to assume that most systems in our universe have already gone through the process of quantum relaxation. However, it is possible that quantum nonequilibrium (or violations of the Born rule) have existed in the very early universe \cite{bib:Valentini1991,bib:Valentini1992,bib:Valentini1996, bib:Valentini2001, bib:Valentini2019, bib:Valentini2010,bib:Valentini2009}.

According to standard inflationary cosmology , the temperature anisotropies observed on the cosmic microwave background (CMB) are generated by quantum vacuum fluctuations of the scalar field \cite{bib:Peter2009}. If some modes of this field were out of equilibrium before the inflationary period it is possible that they would transfer a nonequilibrium signature to the power spectrum of the CMB \cite{bib:Valentini2010,bib:Valentini2007,bib:Valentini2008,bib:Colin2013,bib:Colin2015,bib:Valentini2015,bib:Colin2016, bib:Vitenti2019}. In another cosmological scenario, nonequilibrium could be transferred from some modes of the inflaton to other field modes during reheating, opening up the possibility that violations of the Born rule could have an effect on other stages of the evolution of the early universe (such as particle production or structure formation) or survive in relic systems that could be detected today \cite{bib:Valentini2007,bib:Underwood2015,bib:Underwood2017}. It has also been proposed that black-hole formation and evaporation could lead to the generation of quantum nonequilibrium in the resulting Hawking radiation \cite{bib:Valentini2007,bib:Valentini2004ep,bib:Kandhadai2020, bib:Valentini2021}.  All of those scenarios, however, depend on a deeper understanding of both the quantum relaxation process and the role of quantum mechanics in extreme gravitational situations.

The pilot-wave theory has also been applied successfully to cosmology providing a scenario where the initial singularity can be avoided due to quantum effects in a bouncing model (see \cite{bib:Pinto-Neto2021} for a recent review). The theory has also been used to describe the transition of quantum to classical perturbations during the early stages of the universe both in bouncing and inflationary models \cite{bib:Pinto-Neto2012, bib:Pinto-Neto2013, bib:Pinto-Neto2016, bib:Pinto-Neto2018}. In those applications it was always assumed that the wave function of the universe and all its subsystems were initially in quantum equilibrium but it is also possible that violations of the Born rule could produce detectable effects in these scenarios. In bouncing models, one would have to study the evolution of quantum nonequilibrium of the vacuum perturbations before and during the bounce to observe if they could lead to detectable signatures on the CMB spectrum.  It is also possible that, even if quantum relaxation has taken place in the contracting phase, deviations from equilibrium could be generated during the bounce by novel quantum gravitational processes \cite{bib:Valentini2021}.

In de Broglie-Bohm theory a system is described by an evolving wave function $\psi(q,t)$ that satisfies the usual Schrödinger equation $\imath\partial\psi/\partial t = \mathit{H}\psi$ and also `guides' an actual  configuration $q(t)$ of the system by de Broglie's dynamical equation for the trajectories:
\begin{equation} \label{eq:1}
    \dot{q}(t) = \frac{j(q,t)}{|\psi(q,t)|^{2}},
\end{equation}
where $j(q,t)$ is the usual Schrödinger current. From the Schrödinger equation we can derive a continuity equation
\begin{equation} \label{eq:2}
    \frac{\partial |\psi(q,t)|^{2}}{\partial t} + \partial_q \cdot j(q,t) = 0
\end{equation}
for the density $|\psi(q,t)|^2 $, which has no intrinsic relation with probabilities. A wave function guides the configuration of an \emph{individual} system. However, in actual quantum experiments we consider an ensemble of systems guided by the same wave function but with different initial configurations (or positions, in the case of a particle system). This ensemble has a probability distribution that evolves according to
\begin{equation} \label{eq:3}
    \frac{\partial\rho(q,t)}{\partial t} + \partial_q \cdot (\rho(q,t)\dot{q}(t)) = 0.
\end{equation}
Equations (1.2) and (1.3) take the same form, which means that if initially $\rho(q,t)=|\psi(q, t)|^2 $ this state of `quantum equilibrium' will be preserved for all times $t$. But in the context of de Broglie-Bohm theory we can consider the possibility of nonequilibrium distributions that violate the Born rule. If we explore that possibility, there must be a mechanism through which those nonequilibrium states evolve to the usual distributions that we observe the laboratory.

In classical statistical mechanics the process of relaxation to equilibrium for an initial nonequilibrium probability distribution may be described by the evolution of a coarse-grained \textit{H}-function \cite{bib:Davies1977}. Using the fact that a probability density $p$ and the volume element of phase space $d\Omega$ are conserved along trajectories it can be shown that the classical \textit{H}-function, $H_{class} = \int d\Omega p\ln{p}$, which is essentially the negative of the entropy, is constant in time. However, if we define a coarse-grained density $\bar{p}$ averaged over small cells of the phase space and assume that initially $\bar{p(0)} = p(0)$ it can be shown that the coarse-grained $\bar{H}_{class}$ obeys
\begin{equation}
    \overline{H}_{class}(t)\leq \bar{H}_{class}(0), 
\end{equation}
the classical \textit{H}-theorem \cite{bib:Ehrenfest2014}. In the context of pilot-wave theory, the continuity equations (1.2) and (1.3) imply that the ratio $f = \rho(q,t)/|\psi(q,t)|^2$ is conserved along trajectories. Besides that, analogous to the phase space volume element $d\Omega$, the element $|\psi(q,t)|^2dq$ is also preserved along trajectories in configuration space. It is then possible to define a \textit{subquantum} \textit{H}-function
\begin{equation}
    H = \int |\psi(q,t)|^2 dq f \ln{f}  = \int dq \rho(q,t) \ln{\left( \frac{\rho(q,t)}{|\psi(q,t)|^2}\right)}.
\end{equation}
As the $H_{class}$, this quantity is constant in time. But dividing the configuration space into cells and averaging $\rho$ and $|\psi|^2$ in each cell, one can define the coarse-grained \textit{H}-function
\begin{equation} \label{eq:4}
\overline{H}(t) = \int dq \overline{\rho(q,t)} \ln \left(\frac{\overline{\rho(q,t)}}{\overline{|\psi(q,t)|^2}} \right),
\end{equation}
that can be shown to obey a `subquantum \textit{H}-theorem' 
\begin{equation}
    \overline{H}(t) \leq \overline{H}(0)
\end{equation}
if the initial state has no fine-grained micro-structure ($\overline{\rho(q,0)}=\rho(q,0)$ and $\overline{|\psi(q,0)|^2} = |\psi(q,0)|^2$) \cite{bib:Valentini1990, bib:Valentini1991, bib:Valentini1992, bib:Valentini2005, bib:Valentini2019}. Quantum equilibrium is reached when $\overline{H}(t) = 0$. Whether or not equilibrium is actually reached depends on the characteristics of the initial quantum state, as does the relaxation timescale.

 The expected quantum relaxation has been broadly confirmed by numerical simulations for a number of systems, though there are some caveats. Two classes of  two-dimensional systems (the harmonic oscillator and a particle in a box) have been extensively studied numerically for different superpositions of energy modes of their quantum states. Rapid relaxation, with an approximately-exponential decay of the coarse-grained \textit{H}-function (1.6), has been observed for most cases studied so far \cite{bib:Valentini2005, bib:Efthymiopoulos2006, bib:Colin2010, bib:Towler2012, bib:Efthymiopoulos2017}, with some exceptions for certain states with a small number of superposed modes  (for which sometimes a small nonequilibrium `residue' is found) \cite{bib:Abraham2014}. Quantum relaxation for  a free scalar field on expanding space has also been studied numerically, showing efficient relaxation at short wavelengths and a certain suppression of relaxation at long (super-Hubble) wavelengths \cite{bib:Valentini2008, bib:Colin2013}. More recently, a system of two one-dimensional harmonic oscillators with a linear time-dependent coupling was also studied and the results indicated an approximate exponential decay to equilibrium but further simulations for interacting systems are still needed \cite{bib:Lustosa2020}. 

Considering that the early universe is a prime candidate for the search for nonequilibrium and its detectable effects, particular attention has been paid in recent years to scalar fields initially out of equilibrium evolving on expanding backgrounds, with a view to predicting possible anomalies in the cosmic microwave background (CMB) caused by early violations of the Born rule \cite{bib:Colin2013, bib:Colin2015, bib:Colin2016, bib:Vitenti2019}.  A preliminary prediction for a large-scale power deficit was obtained from a simplified model of a pre-inflationary era \cite{bib:Colin2015, bib:Colin2016}, and a detailed comparison with cosmological data has been made (with some intriguing hints but inconclusive results) \cite{bib:Vitenti2019}. Studies of more complete cosmological scenarios with early nonequilibrium, as well as more detailed predictions of the effects on the CMB, are still needed.

In all these numerical simulations a number of factors can affect the relaxation timescales obtained in each study.  These include the number of energy modes that are superposed, the relative phases in the initial wave function, and the value of the coarse-graining length \cite{bib:Towler2012}. Underwood \cite{bib:Underwood2018} has also considered cases of 'extreme' nonequilibrium, in which the density $\rho$ is initially concentrated in the tails of the wave function where $|\psi(q,t)|^2$ is very small, for which the likelihood of relaxation seems to depend on the total 'vorticity' of $\psi$.

In a recent paper, one of the authors studied the effect of interactions for a time-dependent coupling of two harmonic oscillators \cite{bib:Lustosa2020}. However, this study had several numerical limitations due to the complexity of the wave function that was given in terms of Bessel functions. Furthermore, as it was assumed that the coupling increases linearly in time, it was not possible to separate the effects of the coupling growth and the time-dependence. For these reasons, in this work we take a step back, and focus exclusively on the possible effects that the interaction has on the quantum relaxation process. We will study a system composed of two one-dimensional harmonic oscillators with a linear time-independent coupling. This example is of cosmological relevance, since in the process of reheating in inflationary models, the energy transfer of the inflaton field to other fields is usually modelled by this type of coupling \cite{bib:Brandenberger2010}. Our analysis will be based on the same numerical method used in \cite{bib:Lustosa2020}. We will focus on the effects of the interaction on the relaxation timescales and on the possible existence of nonequilibrium residues. 

In Section II we describe our physical model and the solution of the Schrödinger equation. In Section III we describe our numerical method and setup, comparing its features with a different method used in previous work. In Section IV we analyse the relaxation process for the system considered, by calculating the evolution of nonequilibrium distributions and tracking the behaviour of trajectories. We also obtain relaxation timescales and nonequilibrium residues from the time evolution of the coarse-grained \textit{H}-function, and we study their dependence on the coupling and on the number of modes superposed in the wave function.
\section{Coupled Harmonic Oscillators} \label{sec:couposcillators}
Harmonic oscillators are often used to model a wide range of systems. In quantum field theory, a  free scalar field mode on Minkowski space can be shown to evolve in an analogous way to a standard harmonic oscillator, while on an expanding background the corresponding oscillator has a time-dependent mass and frequency \cite{bib:Valentini2007}. In the inflationary models mentioned above, if the inflaton was in a state of quantum nonequilibrium before inflation, it is possible that some modes of the scalar field are still out of equilibrium during and after inflation \cite{bib:Valentini2010, bib:Colin2013, bib:Colin2015, bib:Colin2016}. Hence, in the process of reheating, the nonequilibrium of the inflaton may be transferred to the other fields when they interact \cite{bib:Brandenberger2010}. It is also possible  that nonequilibrium is generated at the Planck scale \cite{bib:Valentini2014, bib:Valentini2021} during a cosmological bounce resulting from the collapse of a previous contracting phase \cite{bib:Pinto-Neto2005, bib:Pinto-Neto2012, bib:Pinto-Neto2013, bib:Pinto-Neto2016, bib:Pinto-Neto2018, bib:Pinto-Neto2021}. In any scenario where nonequilibrium survives the first stages of cosmological expansion, it can result in observable effects on the CMB or in the initial stages of particle creation.

In the case of inflation, both within the bouncing scenario or the usual standard cosmological model, the decay of the scalar field happens through interactions with other fields \cite{bib:Peter2009}. In models of multi-fluid inflation, the effect of interaction between two scalar fields is crucial in the description of particle creation \cite{bib:Peter2016}. In previous work, one of the authors considered the effects of a time-dependent interaction on the relaxation process \cite{bib:Lustosa2020}. It was in fact already known that a time-dependent expanding background can directly affect the rate of quantum relaxation \cite{bib:Colin2013}. In this work we focus solely on the effects of interaction in a simple model of coupled one-dimensional oscillators.

This model is described by the following Hamiltonian:
\begin{equation}
H(x_{a}, x_{b}, t) = \frac{p_{a}^{2}}{2m} + \frac{m\omega^{2}x_{a}^{2}}{2} +  \frac{p_{b}^{2}}{2m} +  \frac{m\omega^{2}x_{b}^{2}}{2} + \frac{m k}{2}  x_{a}x_{b}.
\end{equation}  
To decouple equation (2.1) we introduce the standard canonical transformation of coordinates for the positions and momenta;
\begin{align}
\begin{pmatrix}
x_{a} \\
x_{b}
\end{pmatrix} = \sqrt{\frac{1}{2m}}  \begin{pmatrix}
1 & 0 \\
0 & -1
\end{pmatrix}\begin{pmatrix}
x_{1} \\
x_{2}
\end{pmatrix},
\\[10pt]
\begin{pmatrix}
p_{a} \\
p_{b}
\end{pmatrix} = \sqrt{\frac{m}{2}}  \begin{pmatrix}
1 & 0 \\
0 & -1
\end{pmatrix}\begin{pmatrix}
p_{1} \\
p_{2}
\end{pmatrix}.
\label{tc2}
\end{align}

It can be shown that the transformed coordinates obey the same commutation relations as the original ones. Writing the transformed Hamiltonian in terms of the new coordinates $x_r$ ($r=1,2$) we have
\begin{equation}
H(x_{1}, x_{2}, t) = \sum_{r=1}^2H_{r} = \sum_{r=1}^2\left(\frac{p_{r}^{2}}{2} +  \frac{\Omega_{r}^2}{2}x_{r}^{2}\right), 
\end{equation}
where the new frequencies are $\Omega_1 = \sqrt{(\omega^2 + k/2)}$ and $\Omega_2 = \sqrt{(\omega^2 - k/2)}$. Because of the negative sign in $\Omega_2$ the values of $k$ are restricted to the range $[0,2\omega^2]$. After quantisation, we can write the Schrödinger equation as
\begin{equation}
    \imath\frac{\partial \Psi}{\partial t} = \sum_{r=1}^2\left( -\frac{1}{2}\frac{\partial^2}{\partial x_r^2} + \frac{1}{2}\Omega_r^2x_r^2\right)\Psi,
\end{equation} 
where we take $\hbar = 1$. The solution can be written as a superposition of eingenstate solutions of the one-dimensional Schrödinger equations $\imath\partial\psi_r/\partial t = \hat{H}_r\psi_r$. The full superposed solution in terms of the original coordinates is then
\begin{equation}
\Psi(x_{a}, x_{b}, t) = \sum_{n_1, n_2} c_{n_{1}, n_{2}} \psi_{n_1}(x_{1}(x_{a},x_{b}),t) \psi_{n_2}(x_{2}(x_{a},x_{b}),t),
\label{totalpsi}
\end{equation} 
where the constants $c_{n_{1}, n_{2}}$ are taken to be $(1/\sqrt{M})e^{\imath 2\pi\theta_{n_{1}, n_{2}}}$ where $M$ is the number of superposed energy modes and $\theta_{n_{1}, n_{2}}$ are randomly chosen phases. For each value of $M$ we  randomly choose a combination of quantum numbers $n_1$ and $n_2$ (using the same method that was described in \cite{bib:Lustosa2020}). The functions $\psi_{n_r}(x_r,t)$ are the usual solutions for the 1D oscillator written in terms of the Hermite polynomials.

To obtain actual trajectories $x_a(t)$ and $x_b(t)$ we construct guidance equations in the transformed coordinates  $x_1(t)$ and $x_2(t)$, transforming the initial coordinates $(x_a(0), x_b(0)) \rightarrow (x_1(0), x_2(0))$ and solving the de Broglie equations
\begin{equation}
\dot{x_r} = \operatorname{Im}\left( \frac{\partial_r \Psi}{\Psi} \right), 
\end{equation}
to obtain different mappings of $(x_1(0), x_1(0)) \rightarrow (x_1(t), x_2(t))$ for an ensemble of trajectories. Before calculating the coarse-grained $\bar{\rho}$, $\overline{|\psi|^2}$ and $\overline{H}(t)$ we transform back to the original coordinates $(x_a(t), x_b(t))$, thus obtaining the evolution of a chosen initial nonequilibrium distribution $\rho(x_a, x_b, t) \neq |\Psi(x_a, x_b, t)|^2$.

\section{Numerical Setup}
\label{sec:timescales}
\subsection{Numerical Methods; backtracking and the forward trajectories method}

The quantum relaxation process depends on a number of  parameters affecting the evolution of the quantum systems and of their probability distributions. In order to observe how the approach to equilibrium changes with those parameters, different methods of analysis have been developed. 

Firstly, for a one-dimensional system of a particle in a box \cite{bib:Valentini2001} an approximate and partial approach to equilibrium was observed on a coarse-grained level. Improved numerical simulations for a two-dimensional system showed a good fit of the data to an \textit{H}-function of the form $\overline{H_0}e^{-t/\tau}$ \cite{bib:Valentini2005, bib:Towler2012}. Furthermore, the effects of the number of modes $M$ of the wave function and the coarse-graining length $\epsilon$ on the relaxation timescales were demonstrated. The results showed a directly correlation between the increase of $M$ and the decrease of $\tau$. A more complete study of the evolution of nonequilibrium distributions for a 2D harmonic oscillator for longer relaxation times was done in \cite{bib:Abraham2014}. In that work, the simulation results showed that for lower values of $M$ it was possible that some residual nonequilibrium remained even after a long time. Those simulations also provided a new best fit for the coarse-grained \textit{H}-function of the form $ a e^{-t/\tau} + R$. The case of a free mode of a scalar field on an expanding spatial background and on Minkowski space was also studied \cite{bib:Colin2013}. Rapid relaxation was observed for the the scalar field (with $M = 25$) evolving in flat space-time but, for the time-dependent expanding background, equilibrium was not reached for long-wavelength (super-Hubble) modes of the scalar field.

All of those studies cited above used a similar \emph{backtracking} method (for a detailed description see \cite{bib:Valentini2005}). This method was successful in describing the relaxation process and obtaining correlations between the relaxation timescales $\tau$ and the values of $M$ and $\epsilon$. However, the backtracking method has the disadvantage of taking a large amount of computer time to calculate values of $\overline{H}(t)$ for different values of $t$. For large number of modes, and for ensembles of trajectories, finding the evolution of the coarse-grained \textit{H}-function becomes computationally intractable. 

A way to avoid this problem is to study the character of the trajectories \cite{bib:Abraham2014, bib:Frisk1997, bib:Efthymiopoulos2006, bib:Efthymiopoulos2007, bib:Efthymiopoulos2017, bib:Kandhadai2016} without calculating the evolution of actual complete distributions. Due to the chaotic nature of many of the quantum trajectories, it is possible to see indications of relaxation, or lack thereof, by analysing selected sets of trajectories. If the trajectories explore large regions of the configuration space this indicates that quantum relaxation can occur more efficiently. On the other hand, if they are confined to small regions that is an indication that quantum equilibrium cannot be reached. Although useful to determine general properties of the trajectories, this method does not allow for a more precise quantification of the relaxation process. 

Another way to study the trajectories and quantum relaxation is related to the \emph{vorticity} \cite{bib:Wisniacki2005, bib:Wisniacki2006} of the velocity fields and it was used to study the time evolution of some extreme nonequilibrium distributions in \cite{bib:Underwood2015}. The vorticity is related to the number of modes $M$ as well as to the chosen random phases $\theta_{n_1,n_2}$ but it is not clear how we can use this method in a more general manner to quantify relaxation timescales.

Finally, in recent work another method was used to study a system of two one-dimensional oscillators with a time-dependent coupling (we shall refer to this method as the Forward Trajectories Method, or FTM) \cite{bib:Lustosa2020}. Using this method, one can calculate the trajectories from $t_0$ to a final time, recording the distribution of trajectories at each time $t$ to obtain the \textit{H}-function. One of the reasons that the backtracking method was used in previous work was to avoid the uneven sampling of distributions when some regions of configuration space contain most trajectories and others have close to none. However, with a large enough number of trajectories this problem does not appear relevant to the accurate calculation of $\overline{H}(t)$. For each time $t$ that we record the distribution, the coarse-grained value of $\rho$ is calculated by summing the number of trajectories that are in each cell. It is the total coarse-grained distribution that is relevant to the calculation of relaxation timescales. The backtracking method is highly sensitive to changes in the  sampling within each cell whereas the FTM method gives approximately the same values of $\overline{H}(t)$ for different coarse-graining lengths.

\begin{figure}
\centering
  \includegraphics[scale=0.3]{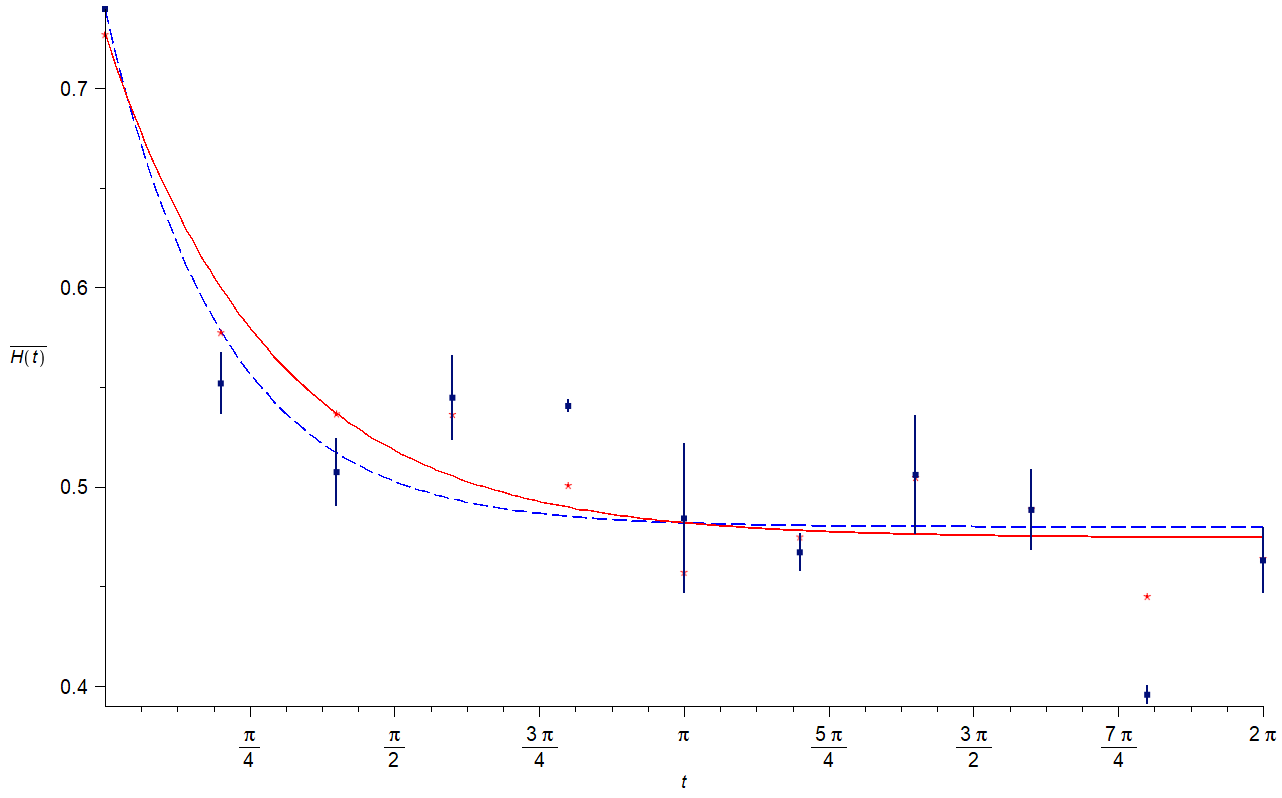}
   \caption{$\overline{H}(t)$ functions calculated using the FTM method (red stars) and the backtracking method (blue stars) The solid lines are the best fit functions of the form  $(\bar{H_0} - R)e^{-t/\tau} + R$ for each data set. }
   \label{backXfrwd}
\end{figure}

Both methods have advantages and limitations, however, the FTM is more efficient and allowed for a wider analysis of the effect of the parameters on the relaxation process for the system considered in \cite{bib:Lustosa2020}. In Figure \ref{backXfrwd} we show a comparison of the two methods for a simple case with $M=9$ for the wave function (2.6). The error bars are obtained from the standard deviation calculated from three different coarse-graining lengths (which leads to different number of points per cell) and the points shown on the graph are the mean values of the \textit{H}-functions. Both cases were calculated on a grid with a total of 256 cells. The two \textit{H}-functions, obtained in the interval $t_0 = 0$ to $t_f=2\pi$, are in relatively good agreement with one another and the bigger spread in some of the points for the backtracked case indicates that the FTM is slightly more precise. The bigger error bars for the backtracking case are a result of the sensitivity of the method to changes in the number of points per cell (the error bars for the FTM are not visible because the values of $\overline{H}(t)$ for different numbers of points per cell are almost the same). Since those changes are practically irrelevant in the method we will use from now on, we will calculate the error bars in a different manner in the next sections. 

\subsection{Our Setup}

The simulations in the next section are all done using the FTM method. Trajectories are calculated using a Runge-Kutta method of order 8 with an initial absolute error tolerance of $10^{-9}$ and then again with a tolerance of $10^{-10}$ \cite{bib:Hairer1993} to solve the set of equations (2.7). If the results of the two calculations have a difference of more then $\delta = 5\times 10^{-3}$ in either coordinate the calculation is repeated with an error tolerance one order of magnitude smaller. This process is repeated until the absolute error tolerance reaches the limit value of $10^{-16}$, if the difference between calculations still exceeds $\delta$ the trajectory is excluded from the distribution. In order to obtain simpler velocity fields we transform the initial positions using equation (2.2) and we transform back to the original coordinates at each time $t$ we wish to calculate the coarse-grained \textit{H}-function. 

The calculation of distributions is done with $N =  230400$ trajectories that evolve in a two-dimensional box in the region $[x_a = -5..5, x_b = -5..5]$. To obtain the coarse-grained functions $\overline{\rho(x_a, x_b,t)}$, $\overline{|\psi(x_a, x_b,t)|^2}$ and $\overline{H}(t)$ we divide the considered region in a grid of $16\times16$ cells. The initial distribution $\rho(x_a, x_b, 0)$ is obtained using a random Gaussian generator for each calculation. As the trajectories evolve some cells will have more points then others and the coarse-grained value of the distribution in each cell is obtained by summing the number of points that are in a given cell at each time $t$.

The wave functions are defined up to factors $c_{n_1, n_2}$. The phases $\theta_{n_1, n_2}$ and quantum numbers $[n_1, n_2]$ are randomly generated for each number of superposed modes $M$. In order to include the random effect of the choice of phases and avoid any bias in the initial wave functions \cite{bib:Underwood2018, bib:Lustosa2020}, we repeat the calculations for different sets of random phases and quantum numbers. Error bars are obtained with the mean deviation calculated from the different values of $\overline{H}(t)$ for different sets of phases and quantum numbers. Finally, all simulations are done with $m = \omega = 1$ and for different values of the coupling constant $k$.

\section{Interaction and the Evolution of Quantum Nonequilibrium}

In this section we analyse how the variation of the value of the coupling constant affects the evolution of an initial nonequilibrium distribution. We will show how the interaction influences the behaviour of trajectories and distributions, as well as affecting the relaxation timescales and nonequilibrium residues. 

\subsection{Distributions and Confinement of Trajectories}

The relaxation process can be analyzed by various methods but it is more easily visualized by the direct observation of the evolution of the nonequilibrium distribution. For a given initial ensemble with density $\rho$, if the system evolves towards equilibrium we expect that the final coarse-grained functions $\overline{\rho}$ and $\overline{|\psi|^2}$ will have approximately the same form \cite{bib:Valentini2005, bib:Towler2012, bib:Abraham2014}. If they are exactly equal, from equation (1.7), we find $\overline{H}(t) = 0$ and the distribution has reached the equilibrium state. However, previous simulations have shown that for a sufficiently low number of superposed modes of the wave function, even after a significantly large time, complete relaxation may not be achieved in some cases \cite{bib:Abraham2014}. Other factors may prevent complete relaxation, such as the form of the initial distribution and the characteristics of the velocity field \cite{bib:Underwood2018}. It has also been shown that a scalar field mode initially out of equilibrium evolving on a time-dependent cosmological background does not relax completely  if the mode wavelength is larger than the Hubble radius  \cite{bib:Colin2013}, indicating that time-dependent Hamiltonians can lead to distributions that avoid relaxation. The time-dependent coupled system studied in \cite{bib:Lustosa2020} also indicated that interactions may delay or even prevent complete relaxation.

\begin{figure}[t]
\centering
\begin{tabular}{c c c}
  \includegraphics[width=42mm, height = 42mm]{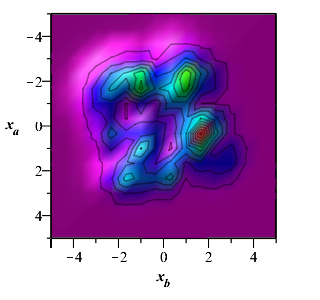} &   \includegraphics[width=42mm, height = 42mm]{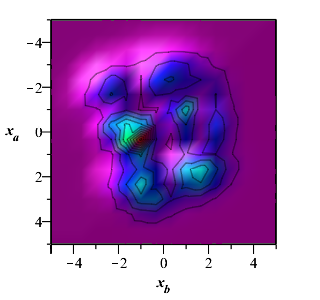} &   \includegraphics[width=42mm, height = 42mm]{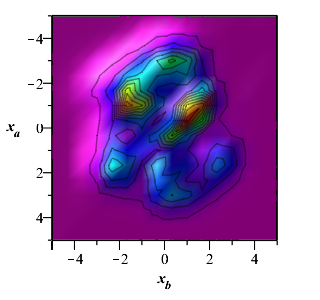} \\
(a) $\overline{|\psi(0)|^2}$  & (b) $\overline{|\psi(5\pi)|^2}$  & (c) $\overline{|\psi(10\pi)|^2}$ \\[6pt]
 \includegraphics[width=42mm, height = 42mm]{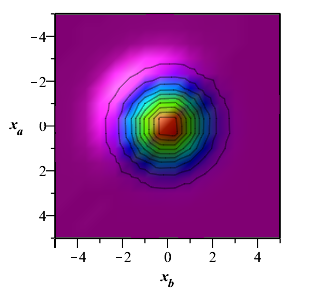} &   \includegraphics[width=42mm, height = 42mm]{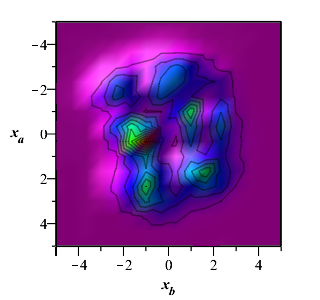} &   \includegraphics[width=42mm, height = 42mm]{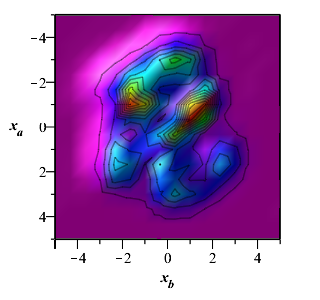} \\
(d) $\overline{\rho(0)}$  & (e) $\overline{\rho(5\pi)}$ & (f) $\overline{\rho(10\pi)}$
\end{tabular}
\caption{Coarse-grained distribution and $|\psi|^2$ at times $t=0$ ((a) and (d)), $t=5\pi$ ((b) and (e)) and $t=10\pi$ ((c) and (f)) for the case of $24$ superposed modes of the wave function (2.6) and coupling constant $k=0.1$. The initial $\overline{\rho}$ is very close to $\overline{|\psi|^2}$ at $t=5\pi$ and is virtually indistinguishable from the equilibrium distribution at $t=10\pi.$}
\end{figure}

\begin{figure}[t]
\centering
\begin{tabular}{c c c}
  \includegraphics[width=42mm, height = 42mm]{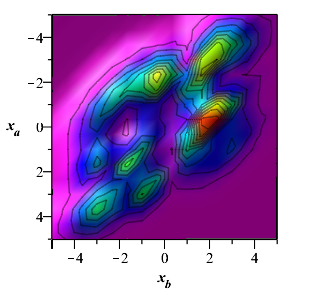} &  
   \includegraphics[width=42mm, height = 42mm]{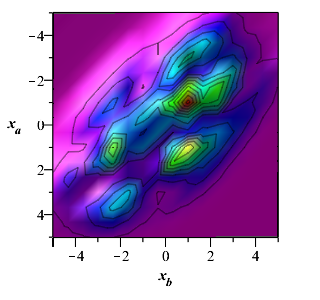} &
   \includegraphics[width=42mm, height = 42mm]{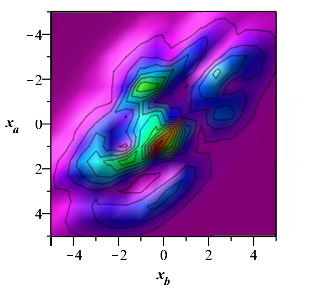} \\
(a) $\overline{|\psi(0)|^2}$  & (b) $\overline{|\psi(5\pi)|^2}$  & (c) $\overline{|\psi(10\pi)|^2}$ \\[6pt]
\includegraphics[width=42mm, height = 42mm]{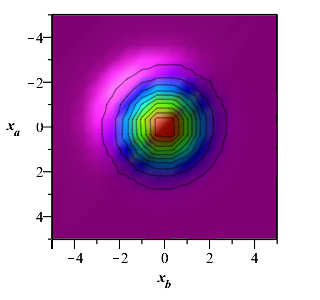} &   \includegraphics[width=42mm, height = 42mm]{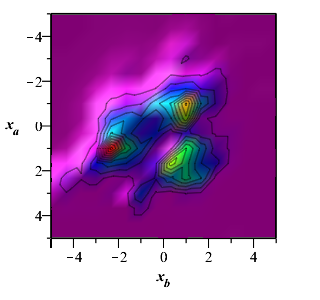} &   \includegraphics[width=42mm, height = 42mm]{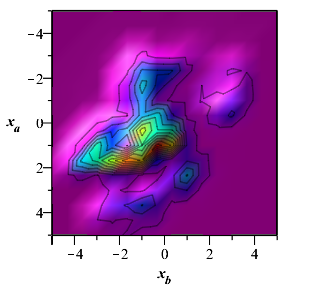} \\
(d)  $\overline{\rho(0)}$  & (e) $\overline{\rho(5\pi)}$ & (f) $\overline{\rho(10\pi)}$
\end{tabular}
\caption{Coarse-grained distribution and $|\psi|^2$ at times $t=0$ ((a) and (d)), $t=5\pi$ ((b) and (e)) and $t=10\pi$ ((c) and (f)) for the case of $24$ superposed modes of the wave function (2.6) and coupling constant  $k=1.8$. The initial coarse-grained nonequilibrium Gaussian distribution does not reach an approximate equilibrium state at $t=10\pi$, indicating that the interaction might delay or prevent the relaxation process.} 
\end{figure}

In this work we focus on how a constant interaction, determined by the value of the coupling constant $k$, affects the evolution of a nonequilibrium distribution. As a preliminary illustration, Figure 2 shows the evolution of an initial coarse-grained nonequilibrium Gaussian distribution and the coarse-grained $|\psi|^2$ given by (2.6) from $t=0$ to $t=10\pi$ for the case with $M=24$ and $k=0.1$. The distribution evolves efficiently to an approximate equilibrium state and is virtually indistinguishable from the $\overline{|\psi|^2}$ at the final time. In contrast, in Figure 3 we show the evolution of $\overline{\rho}$ and $\overline{|\psi|^2}$ with the same number of modes but with $k=1.8$. It is clear that in this case complete relaxation is not achieved at time $t=10\pi$ and $\overline{\rho}$ and $\overline{|\psi|^2}$ differ significantly. 

The evolution of distributions shown in Figures 2 and 3 is a preliminary indication of the effect of interactions on quantum relaxation. To further analyse the relaxation process it is useful to observe how trajectories evolve individually.  Quantum trajectories are often chaotic \cite{bib:Efthymiopoulos2006, bib:Efthymiopoulos2007, bib:Efthymiopoulos2017} and tend to explore most of the support region of the wave function in cases where efficient relaxation takes place. The number of superposed modes of the wave function directly influences the character of the trajectories since they determine the number of nodes of the wave function \cite{bib:Towler2012, bib:Underwood2018}. When trajectories approach those nodes the velocity field becomes rapidly-varying and the trajectories tend to be more chaotic for higher numbers of modes. Chaotic trajectories usually explore a wide region of configuration space and trajectories that are initially very close can have very distant final positions. When most trajectories have those characteristics the approach to equilibrium occurs efficiently. On the other hand, the confinement of trajectories to certain regions is directly related to systems that avoid complete relaxation \cite{bib:Abraham2014}.

To analyse the characteristics of the trajectories for our system of coupled oscillators we make use of two tests that provide a visual representation of how given sets of trajectories explore the configuration space \cite{bib:Abraham2014}. The first test consists of plotting the initial and final positions of five sets of neighbouring trajectories to observe if they spread over the whole support region of the wave function or if they stay confined and close to one another. In the second test we observe the full evolution of 5 individual trajectories to see how they explore the configuration space. In Figure 4 we plot the results of those tests for the case where the wave function has a superposition of 24 modes and for four values of the coupling constant $k$. The right column shows 5 trajectories with randomly selected initial positions\footnote{The initial positions are (2.39,-1.94), (1.27, 3.76), (2.21,-0.01), (-0.89, 2.04) and (-2.30, -0.76).} evolving from $t=0$ to $t=10\pi$. On the left column we have 5 sets of 25 trajectories centered on the initial positions used on the right column plots. For $k=0.1$ and  $k=0.9$ most trajectories are not confined and spread through the bulk of the support region of the wave function. For the case $k=1.1$ the blue trajectories are strongly confined to a region close to their initial positions and most of the green trajectories are also confined. The case with  $k=1.8$ shows confinement to regions close to the initial positions for the purple and blue sets of trajectories and also shows confinement to a certain region for more than half of the black, green and red sets.

These results indicate that for $k < 1$ there is little or no confinement, while for $k > 1$ the confinement becomes increasingly stronger. Simulations with different numbers of modes also corroborate the indication that there might be two regimes  ($k < 1$ and $k > 1$)  for the system we are studying. As we will discuss, this might be explained by the relation between the values of the fundamental frequencies $\omega$ and $k$. Further analysis of this hypothesis will be done in subsection 4(c) by analysing dependence of the relaxation timescales and nonequilibrium residues on the coupling constant.
\begin{figure}[p]
\centering
\begin{tabular}{c c}
  \includegraphics[width=45mm, height = 43mm]{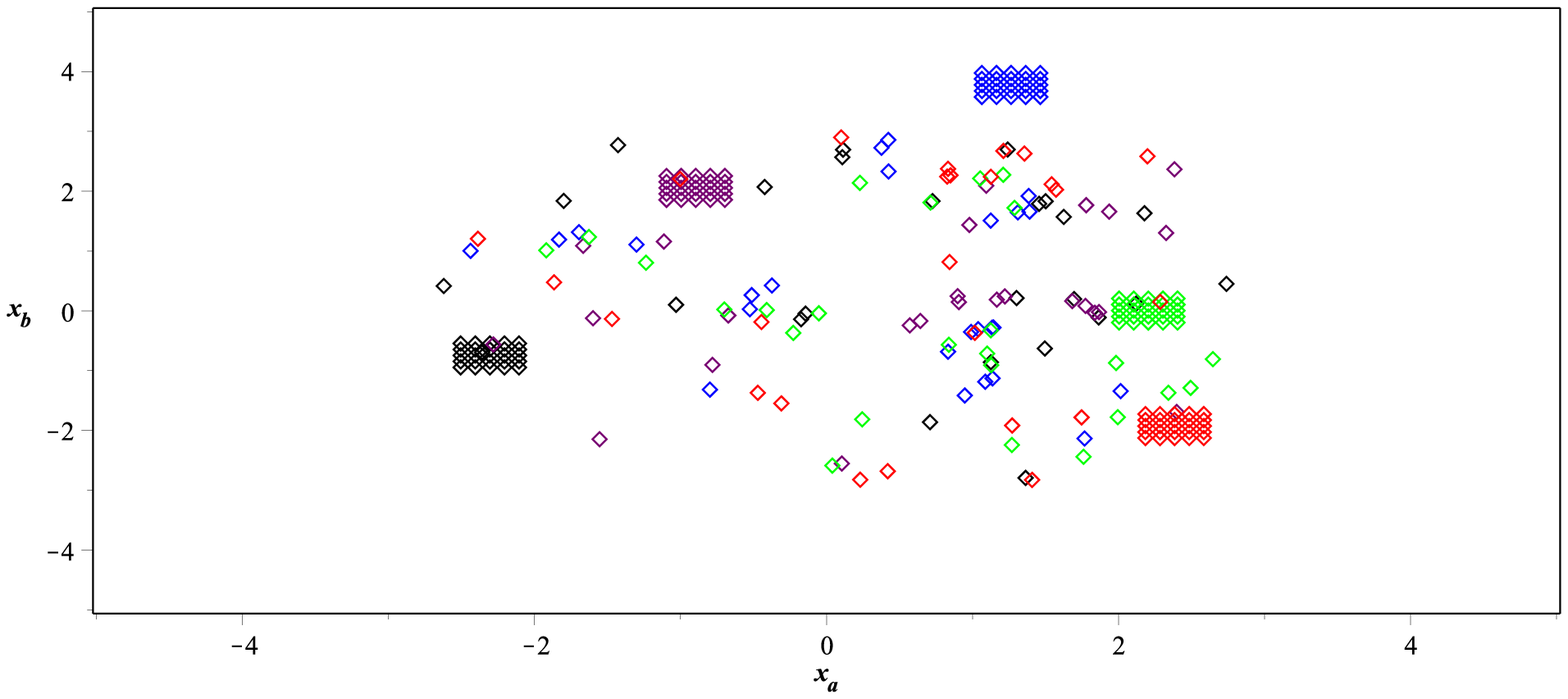} &   \includegraphics[width=45mm, height = 43mm]{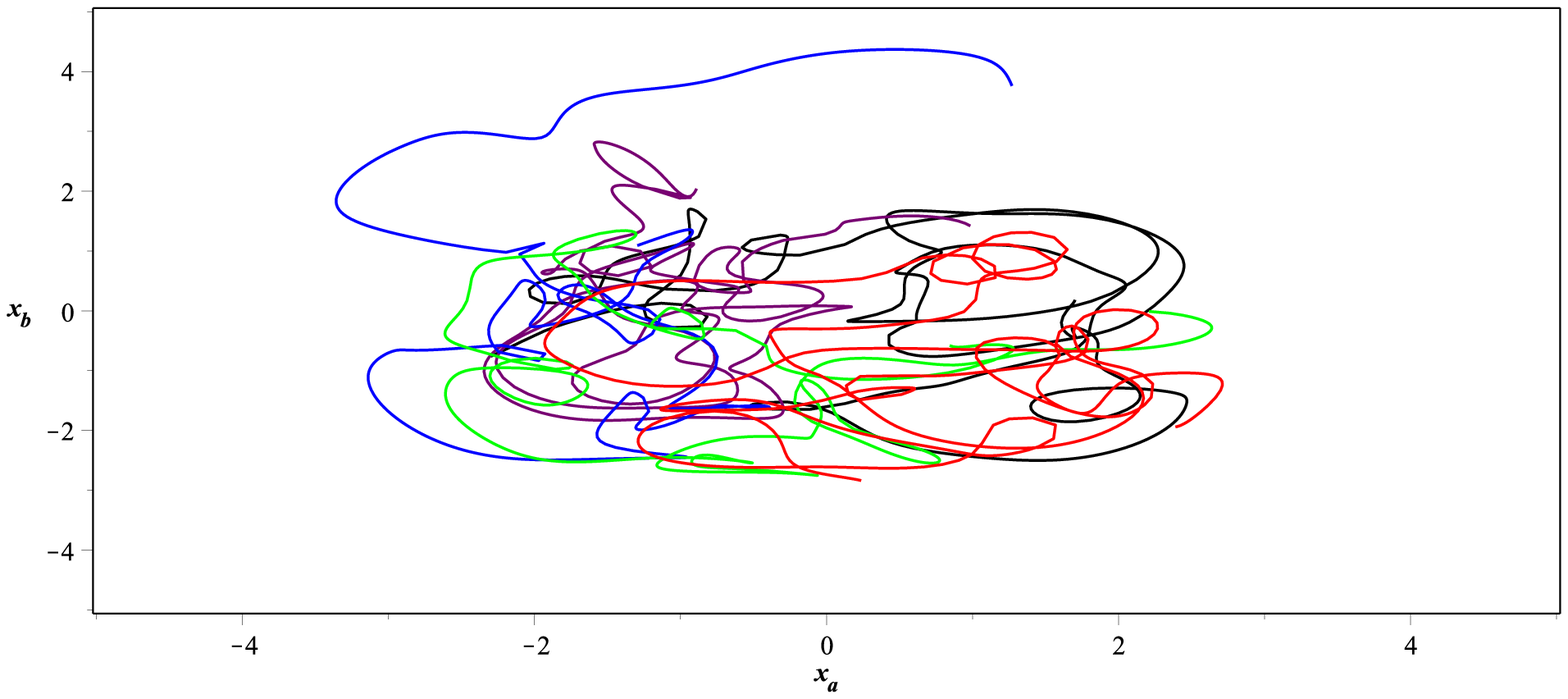} \\
(a)  & (b)  \\[6pt]
 \includegraphics[width=45mm, height = 43mm]{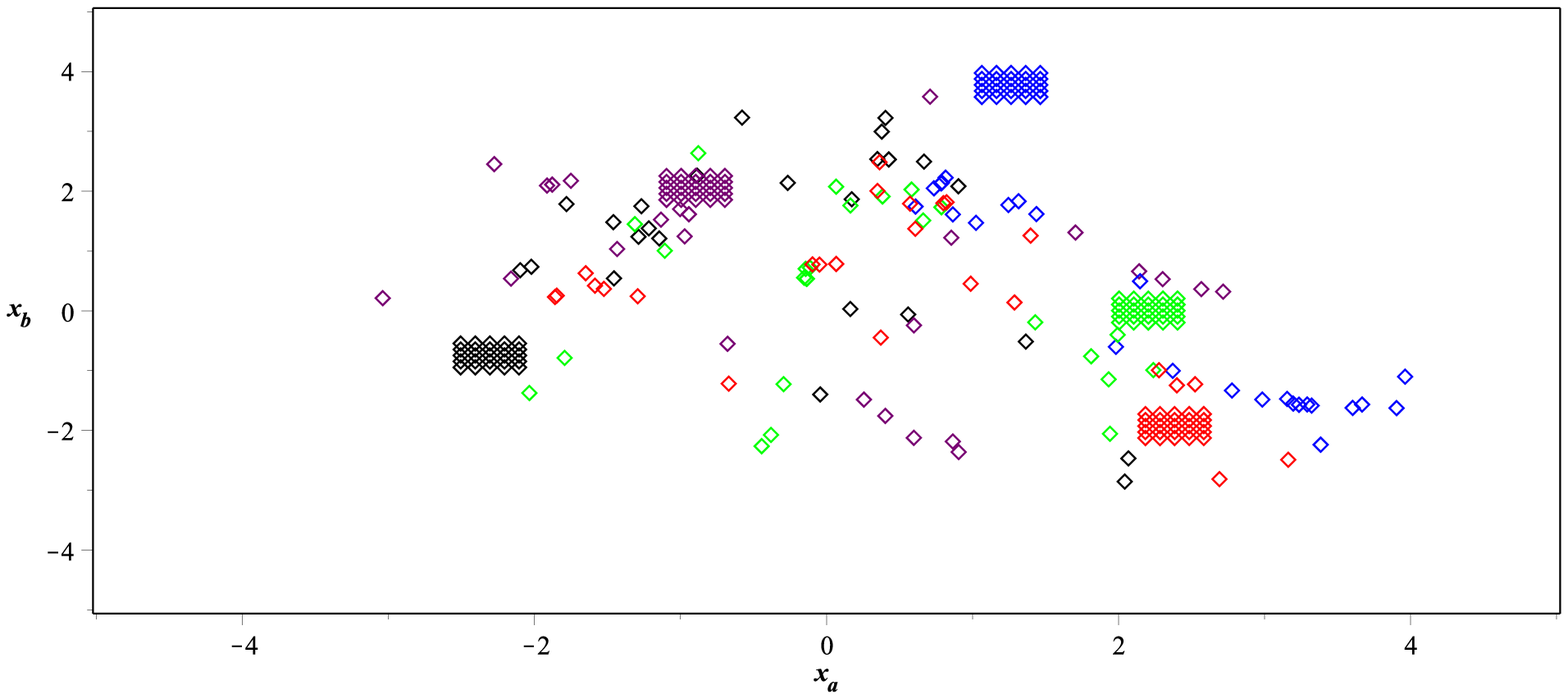} &   \includegraphics[width=45mm, height = 43mm]{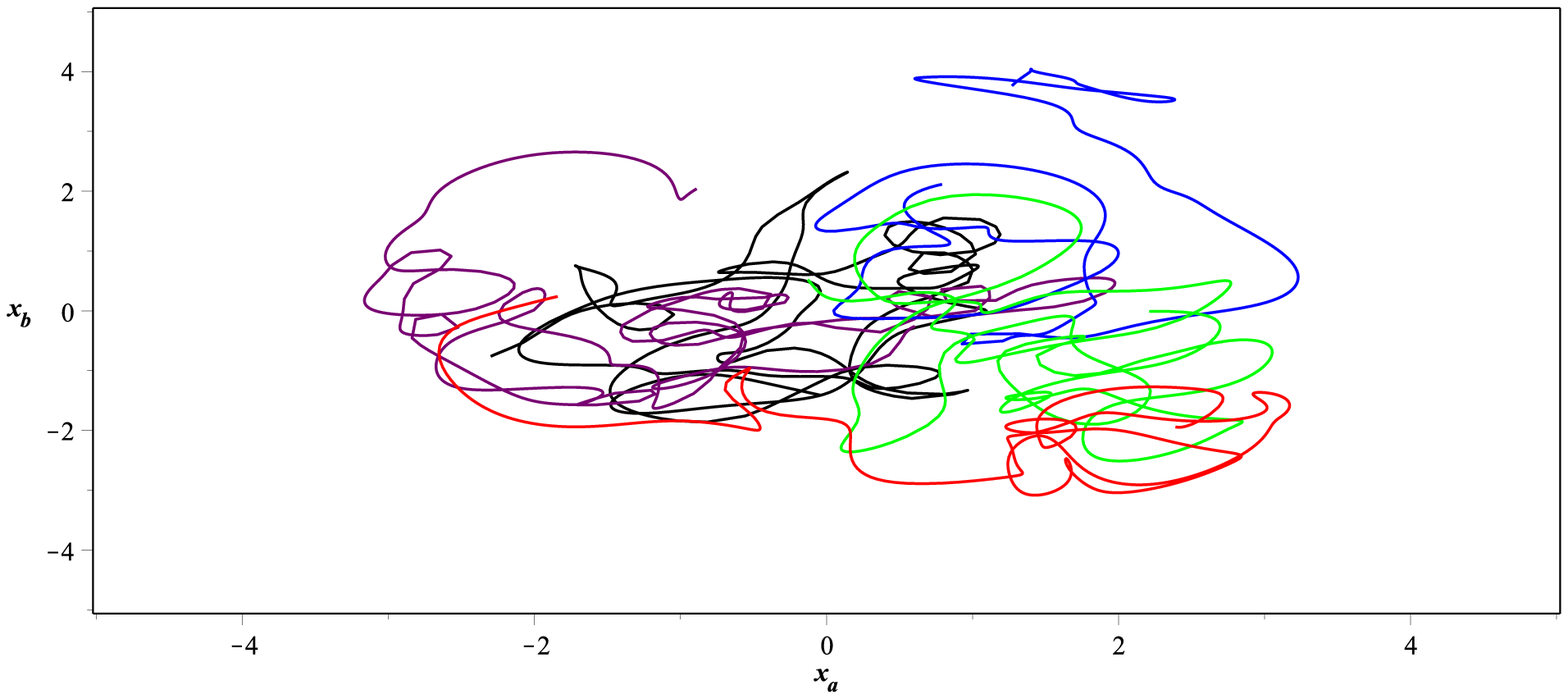} \\
(c)  & (d)\\[6pt]  
\includegraphics[width=45mm, height = 43mm]{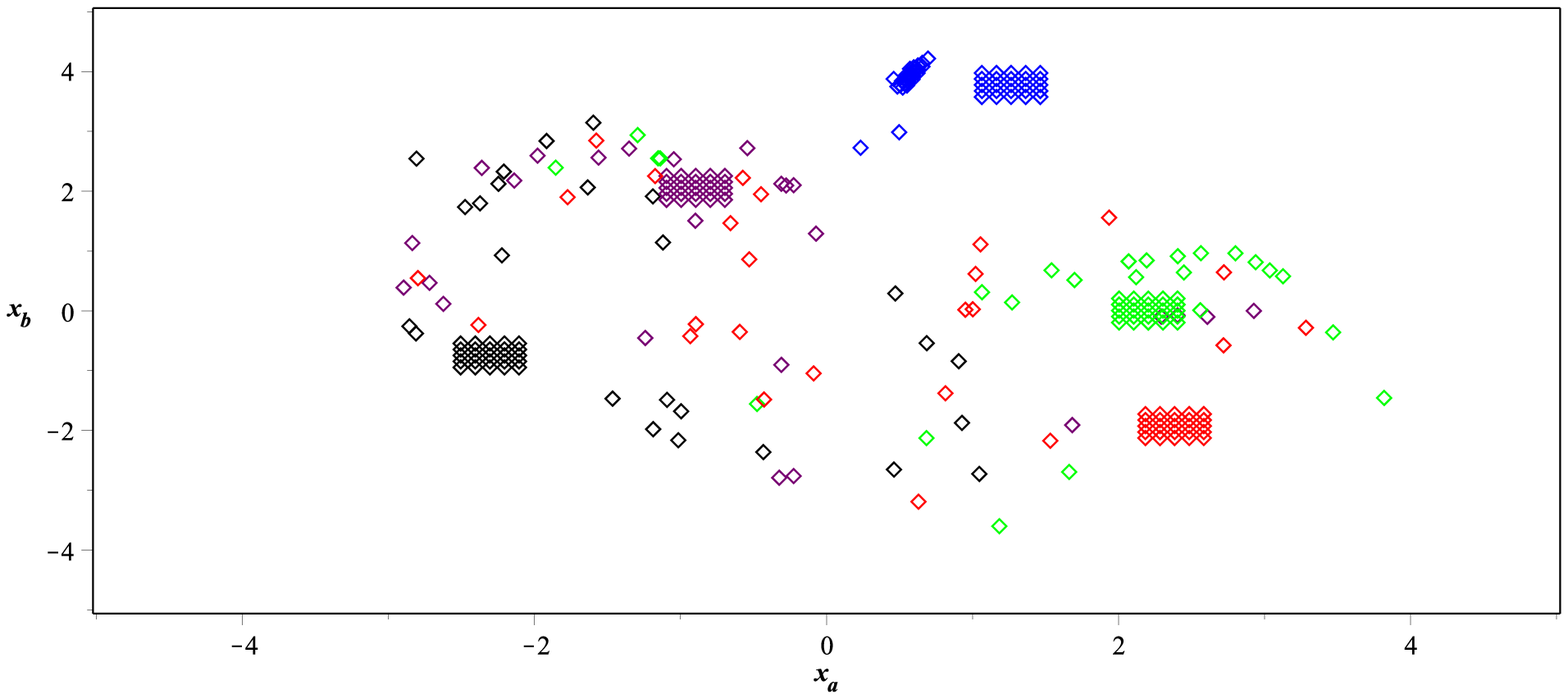} &   \includegraphics[width=45mm, height = 43mm]{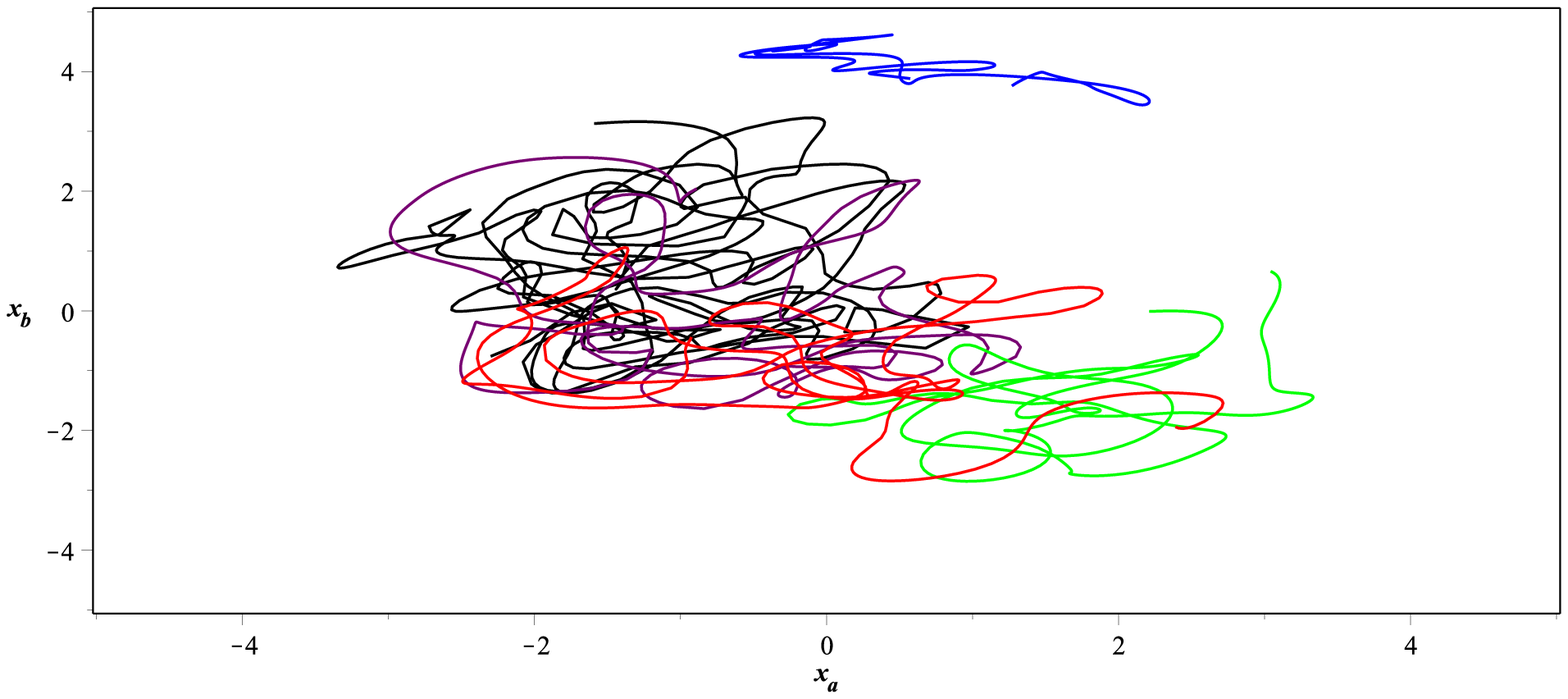} \\
(e)  & (f)  \\[6pt]
 \includegraphics[width=45mm, height = 43mm]{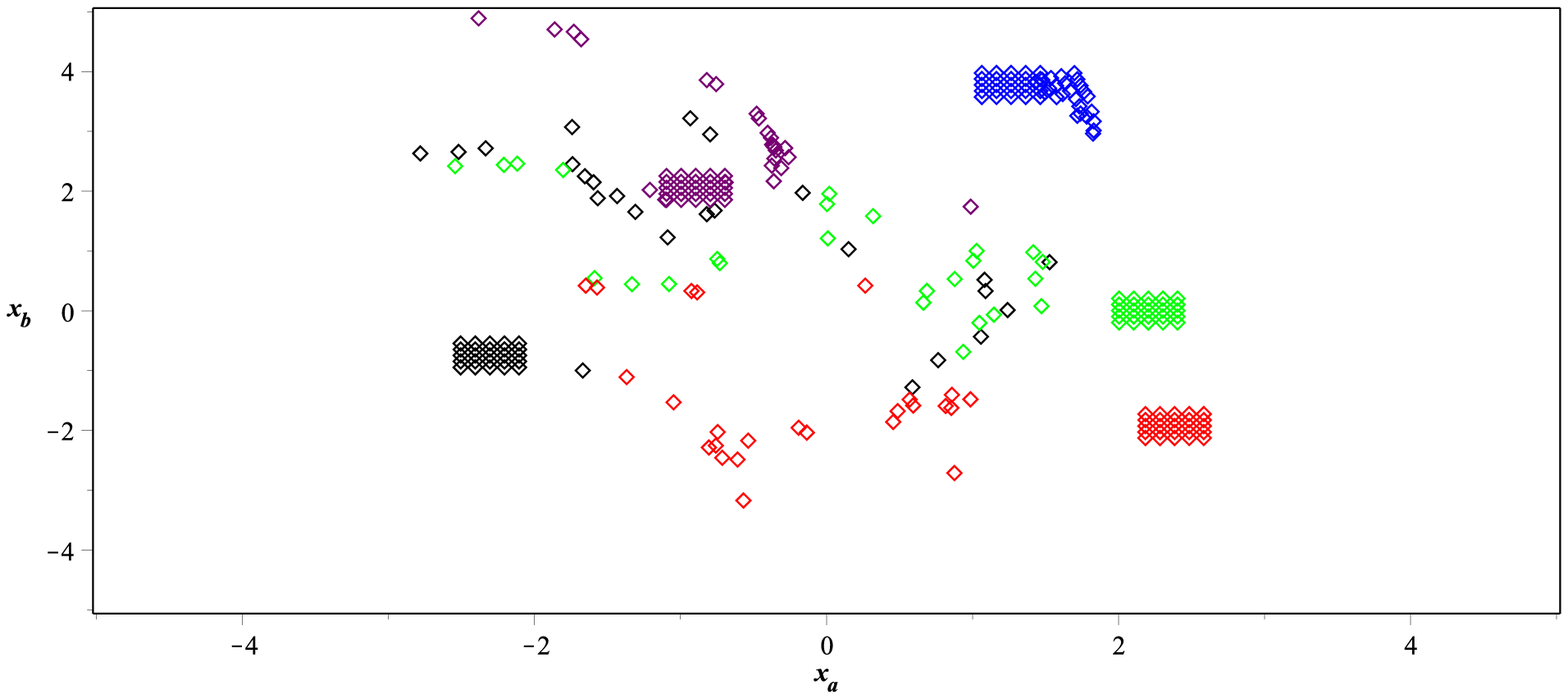} &   \includegraphics[width=45mm, height = 43mm]{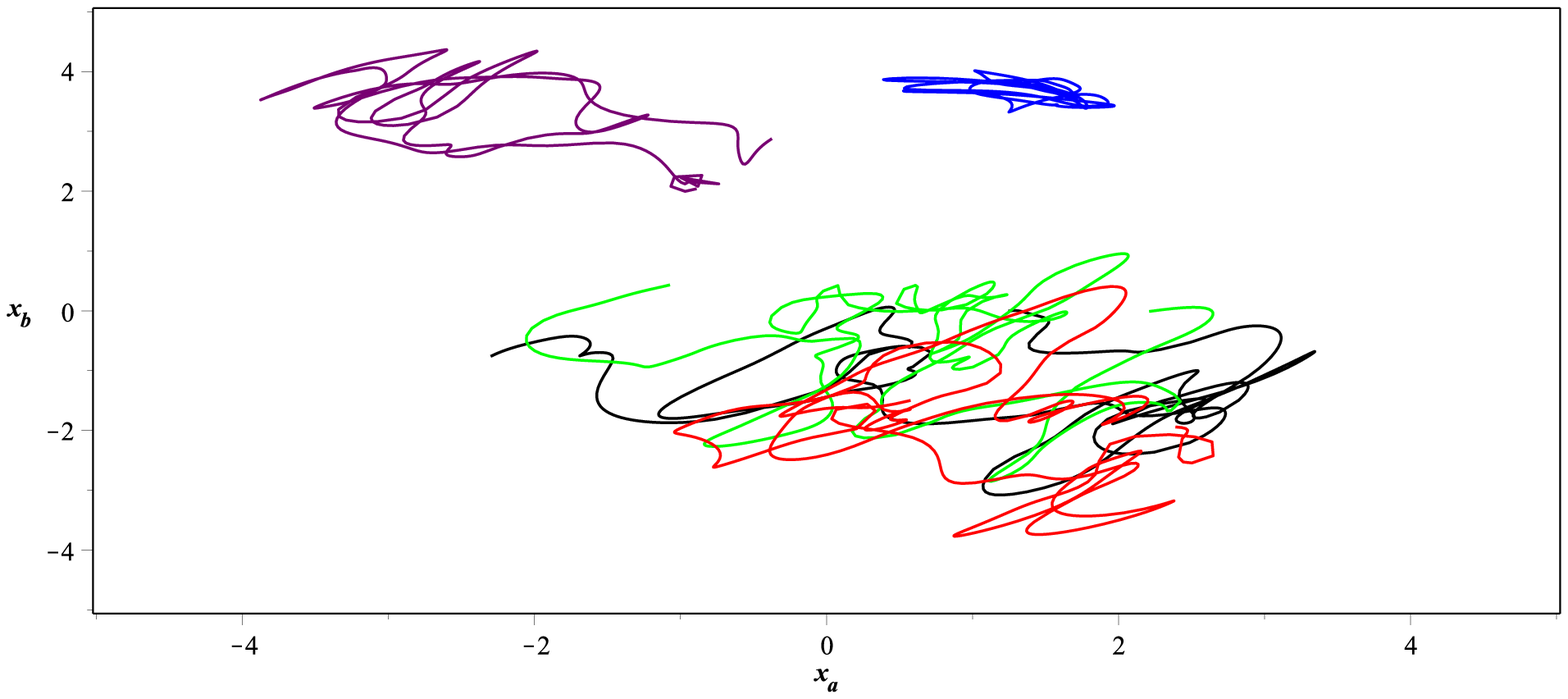} \\
(g)  & (h) 
\end{tabular}
\caption{We display initial and final positions ((a), (c), (e) and (g)) of 5 sets of 25 trajectories with close initial positions. In ((b), (d), (f) and (h)) we show 5 full trajectories. All calculations were done with superposition of $20$ modes on the wave function and for increasing (top to bottom) coupling constant $k = 0.1, 0.9, 1.1$ and $1.8$.}
\end{figure}


\subsection{Long Time Nonequilibrim Residues}

In previous works, numerical simulations that calculate the coarse-grained \textit{H}-function were usually done for relatively short periods of time (except in \cite{bib:Abraham2014}). This can be mostly attributed to the computational difficulties related to calculating a large enough number of trajectories to accurately model a distribution, especially when considering wave functions with larger numbers of modes. However, considering the approximate exponential decay of most \textit{H}-functions calculated so far, those short time simulations are very useful to estimate relaxation timescales $\tau$ for a fitting function of the form $\bar{H_0}e^{-t/\tau}$. 

\begin{figure}[t]
\centering
\begin{tabular}{cc}
  \includegraphics[width=65mm, height = 35mm]{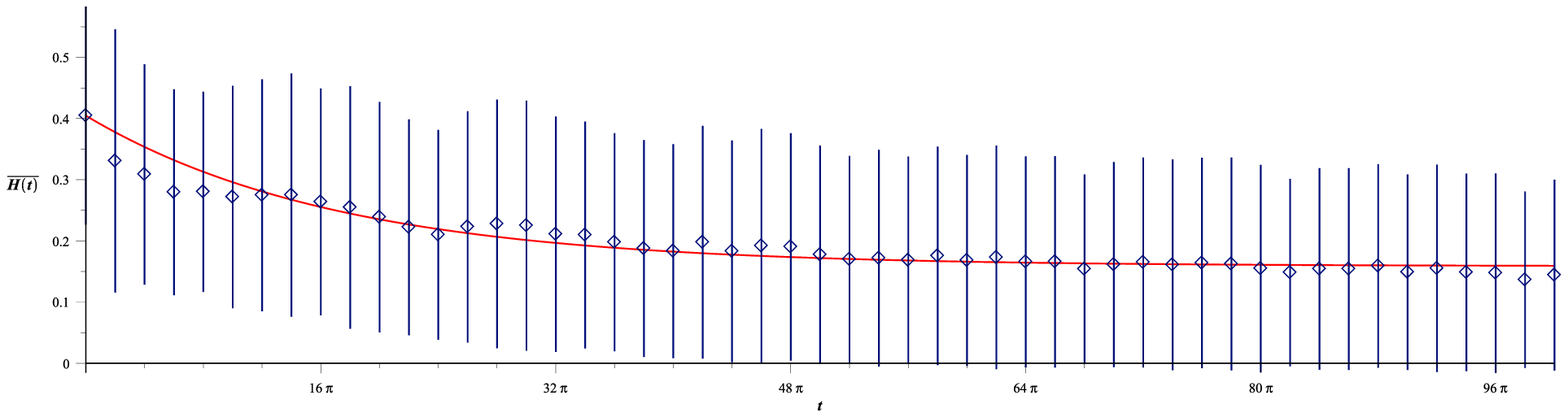} &   \includegraphics[width=65mm, height = 35mm]{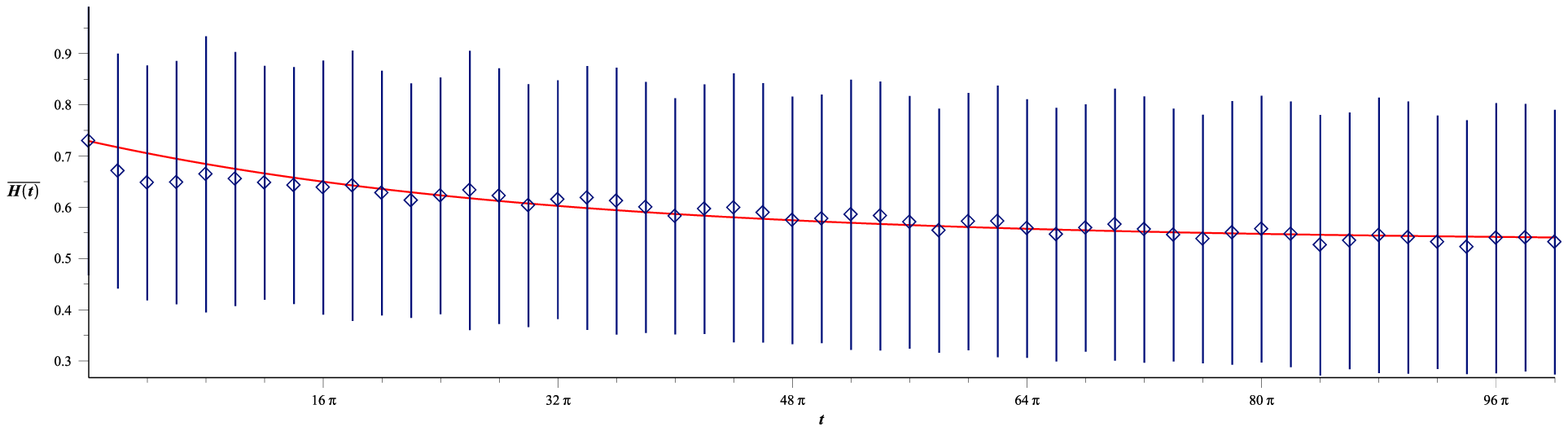} \\
(a) $M = 4$ and $k = 0.5$  & (b) $M = 4$ and $k = 1.8$ \\[6pt]
 \includegraphics[width=65mm, height = 35mm]{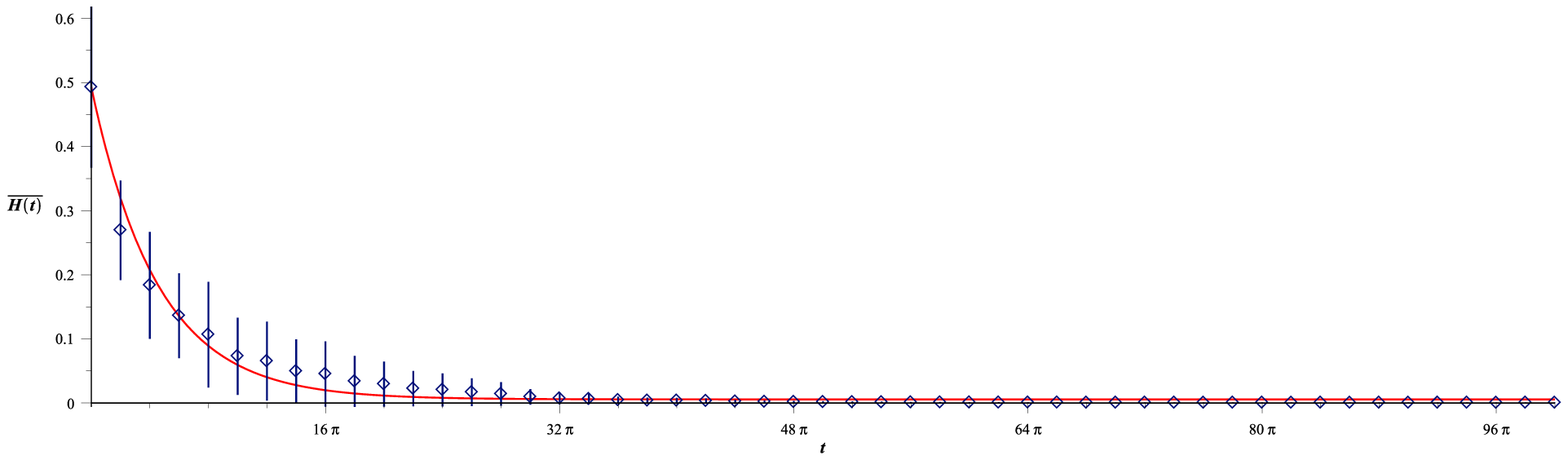} &   \includegraphics[width=65mm, height = 35mm]{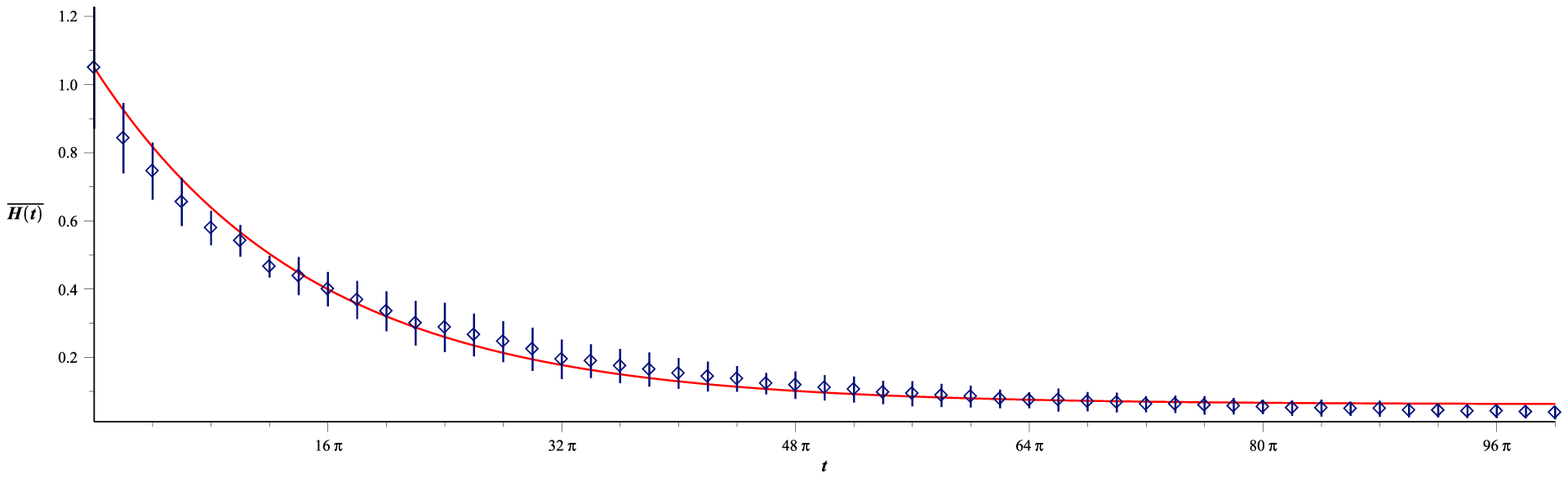} \\
(c) $M = 12$ and $k = 0.5$ & (d) $M = 12$ and $k = 1.8$ \\[6pt]
 \includegraphics[width=65mm, height = 35mm]{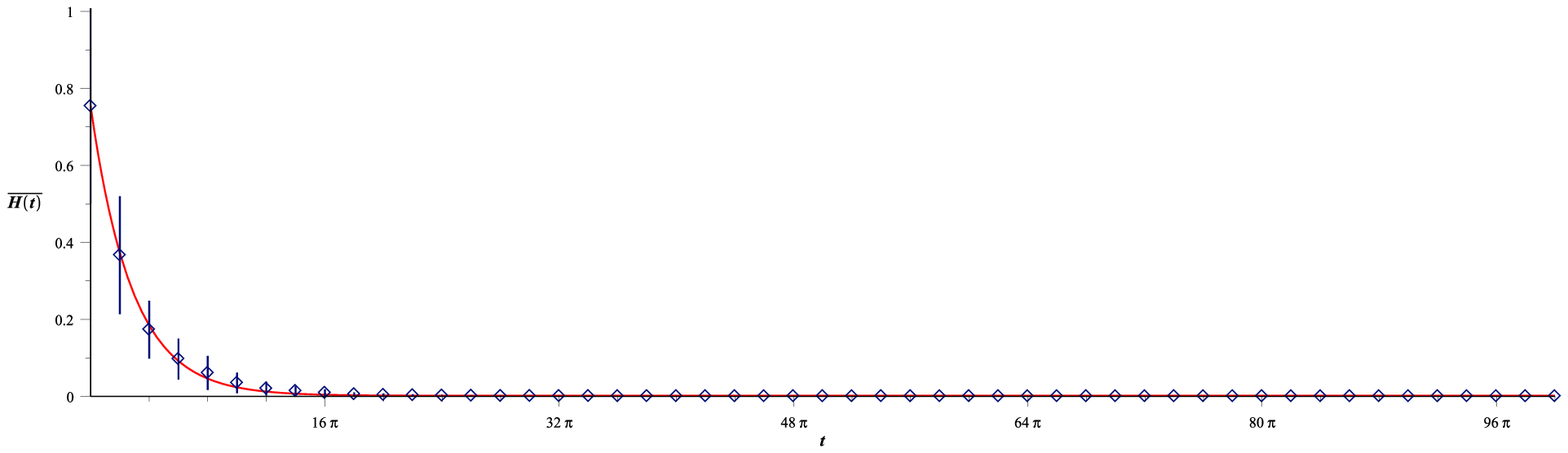} &   \includegraphics[width=65mm, height = 35mm]{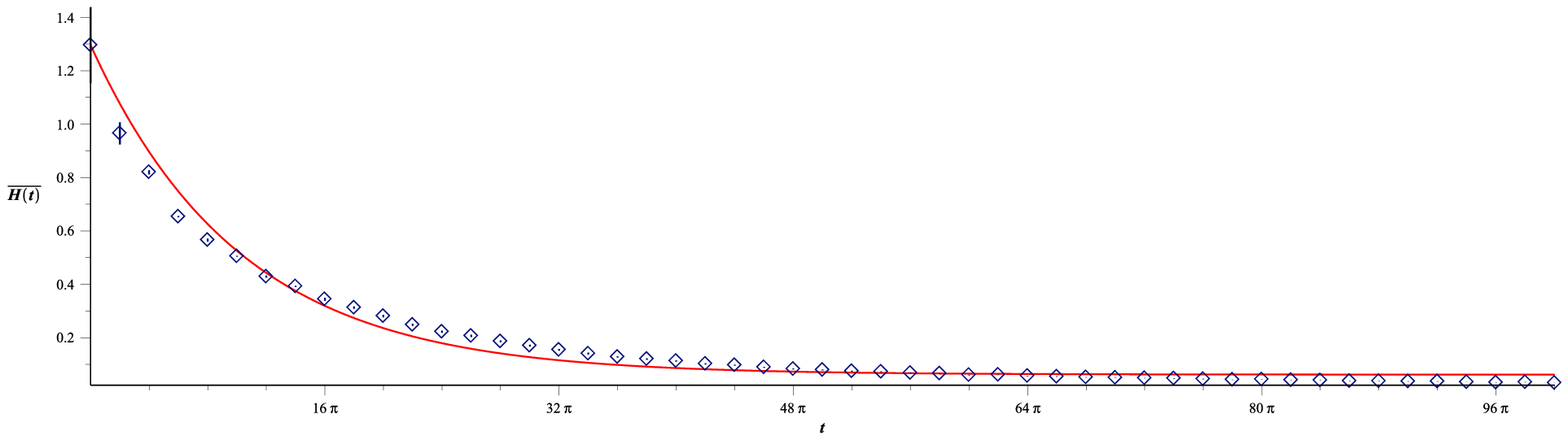} \\
(e) $M = 20$ and $k = 0.5$  & (f) $M = 20$ and $k = 1.8$
\end{tabular}
\caption{$\overline{H}(t)$ functions averaged over three presets of random phases $\theta_{n_1,n_2}$ calculated in the time interval  $[0, 100\pi]$. For $k = 0.5$ there is complete relaxation at an efficient rate for the cases $M=12$ and $M=20$. When $k = 1.8$ relaxation timescales are considerably larger and there are some nonequilibrium residues in all cases. The red line represents the best fit function $(\bar{H_0} - R)e^{-t/\tau} + R$.}
\end{figure}

Longer time simulations are needed to verify if, after the initial decay of the \textit{H}-function, complete relaxation is actually achieved (within the precision limits of the numerical simulations). If that is the case, the value of $\overline{H}(t)$ becomes approximately zero and the system is considered to be in complete equilibrium. Another possibility, which has been observed for some cases with a small number of modes \cite{bib:Abraham2014}, is that a residual nonequilibrium remains even after a long time, indicating that the Born rule might cause observable effects for such systems. Systems that present this feature can then plausibly model realistic situations where quantum nonequilibrium might be detected.

In Figure 5 we show examples of the evolution of \textit{H}-functions for a time interval of $100\pi$.\footnote{This value is given in the units $\hbar=c=1$.} The calculations of $\overline{H}(t)$ are done for three different sets of randomly-chosen initial phases $\theta_{n_{1}, n_{2}}$ and the results are then averaged. The error bars are given by the standard deviation of the three values of the \textit{H}-function (at each time $t$). The red lines represent the best fit of the form $(\bar{H_0} - R)e^{-t/\tau} + R$. Values for $\tau$ and $R$ are obtained by fitting each \textit{H}-function result and then averaging over the three relaxation timescales and residues obtained from the different sets of initial phases.

The case with $M = 4$ decays very slightly for both values of $k$ but the average value of $\tau$ is considerably higher for the case with higher coupling constant ($\tau \approx 56.43$ for $ k = 0.5$ and $\tau \approx 101.65$ for $ k = 1.8$). The averaged residues also represent a higher percentage of $\overline{H}(0)$ for the case with stronger interaction ($R \approx 38.13\%$ for $ k = 0.5$ and $R \approx 72.48\%$ for $ k = 1.8$).

The cases with $M=12$ and $M=20$ for the coupling constant $k=0.5$ show approximately complete relaxation. The average residues for both cases represent approximately $0.5\%$ of the initial value of the coarse-grained \textit{H}-function. The relaxation timescale is smaller for the case with a higher number of modes. The cases with $k=1.8$ show a considerable increase both in the values of $\tau$ and $R$. However, for a sufficiently long time both cases have a nonequilibrium residue of approximately $5\%$ of the initial value of $\overline{H}(t)$, which cannot be considered negligible. These results show that a  larger coupling can significantly delay, and even somewhat suppress, quantum relaxation (especially for a sufficiently small number of modes). This might allow for nonequilibrium effects to be detectable in certain situations. In the next subsection we will further explore the relation between $\tau$, $R$ and $k$. 

\begin{figure}[t]
\centering
\begin{tabular}{cc}
  \includegraphics[width=65mm, height = 35mm]{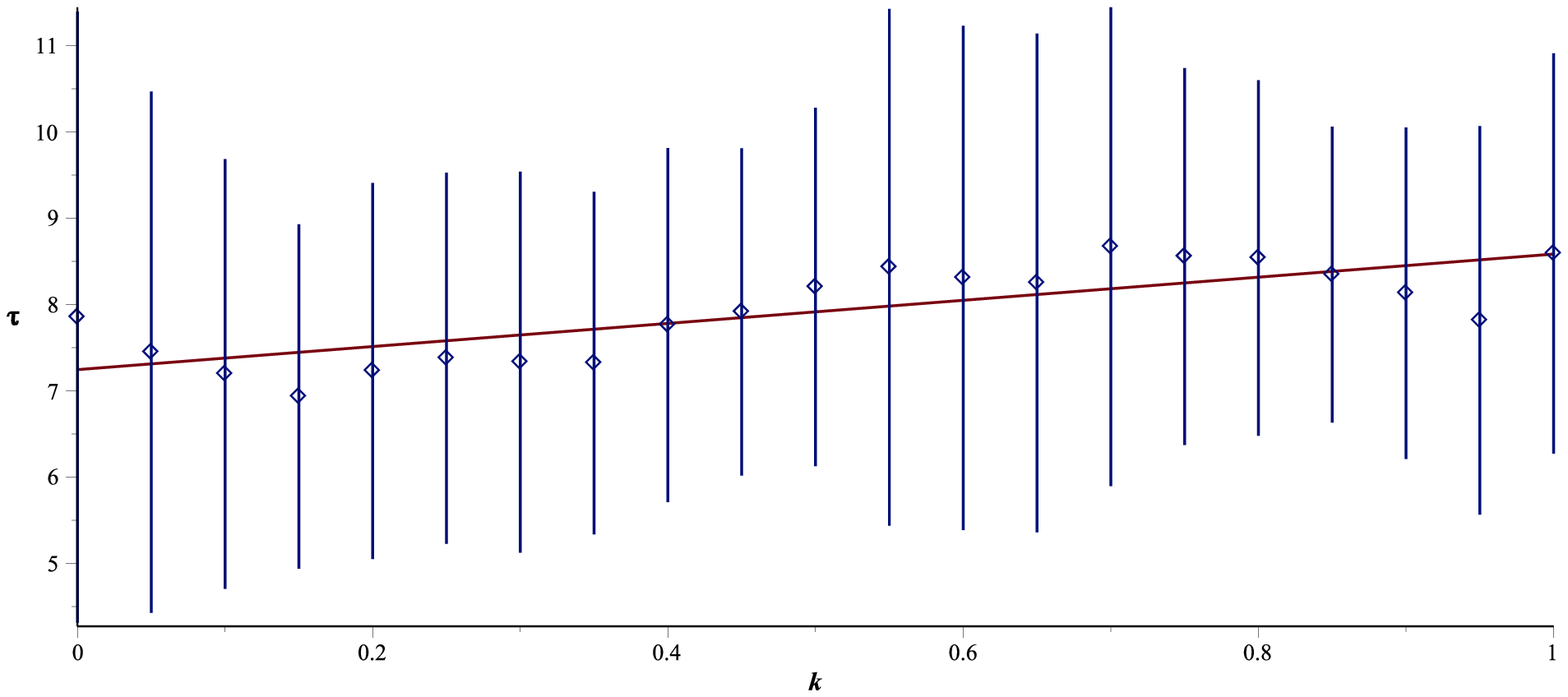} &   \includegraphics[width=65mm, height = 35mm]{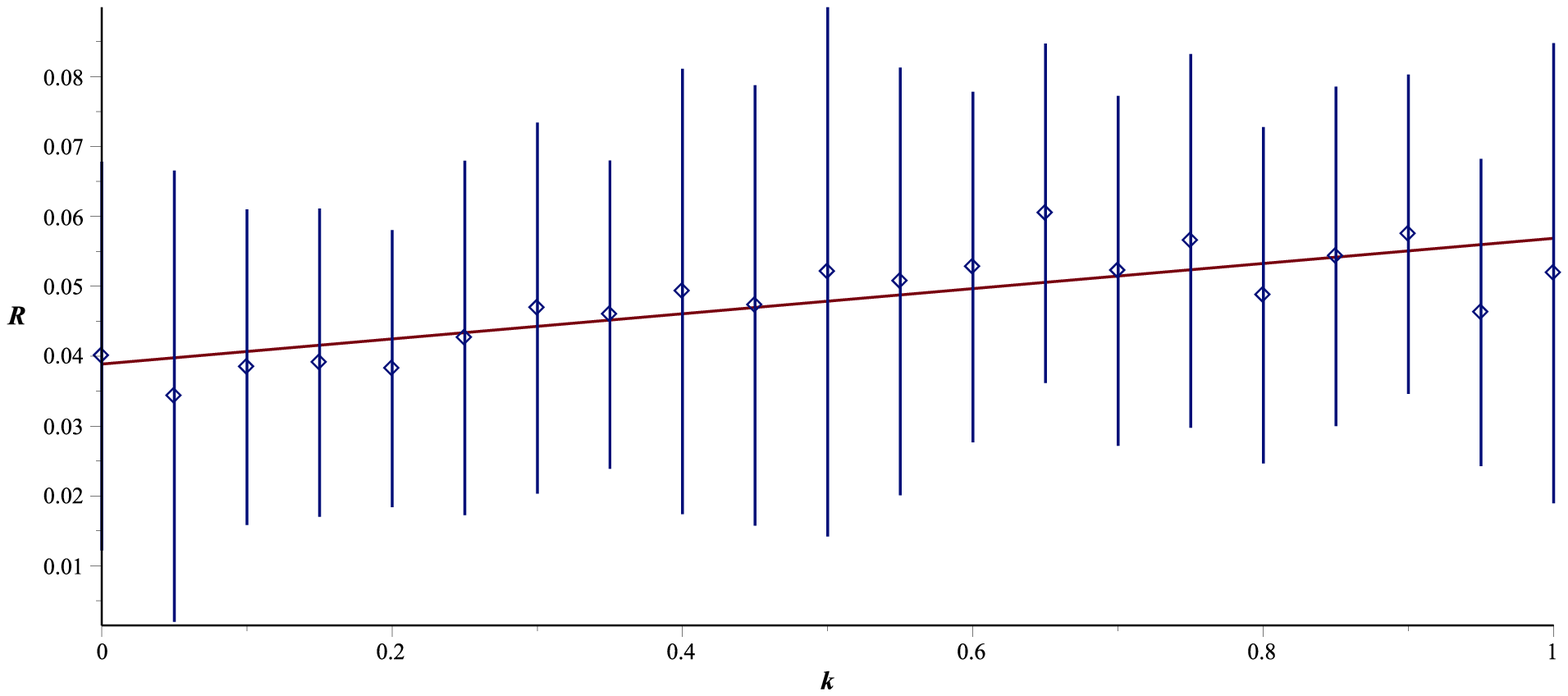} \\
(a)$\tau(k)$ for $M = 12$. & (b)$R(k)$ for $M = 12$.  \\[6pt]
 \includegraphics[width=65mm, height = 35mm]{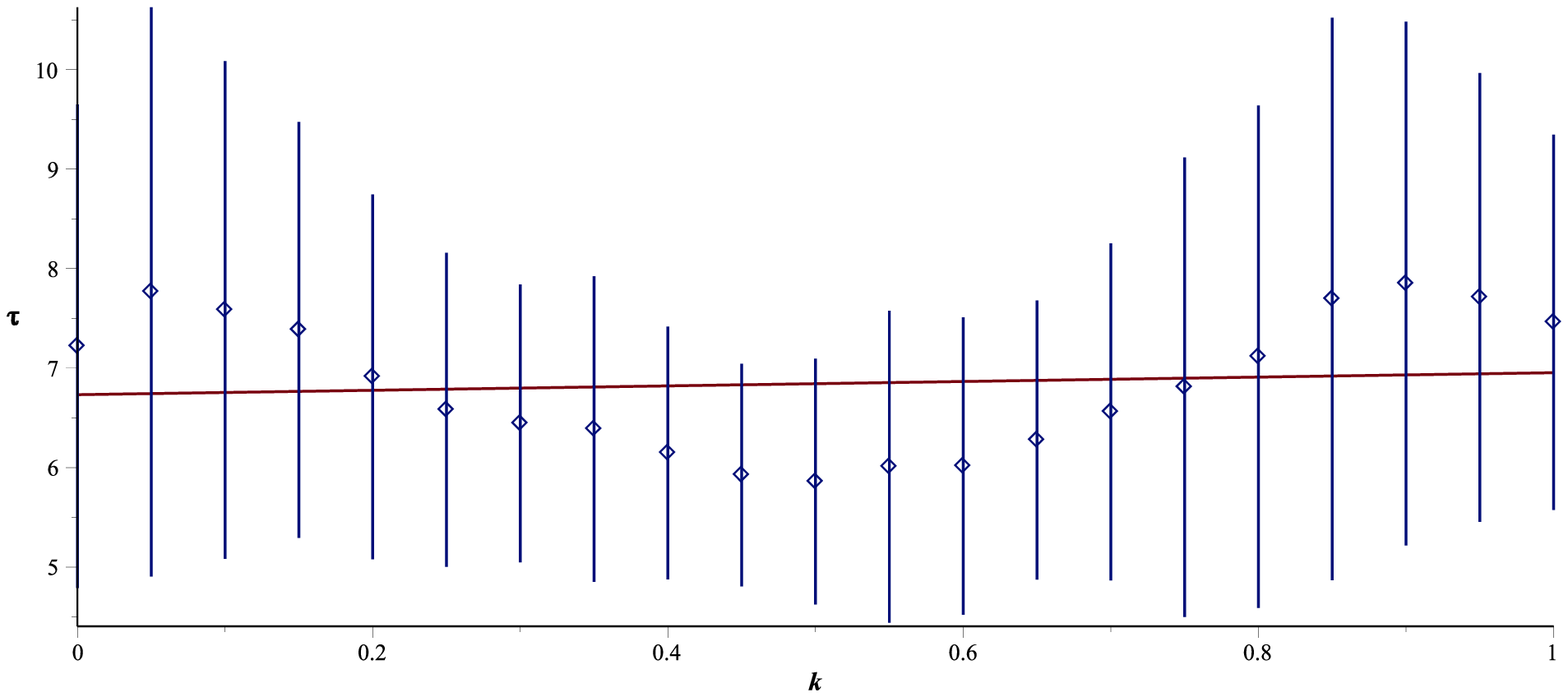} &   \includegraphics[width=65mm, height = 35mm]{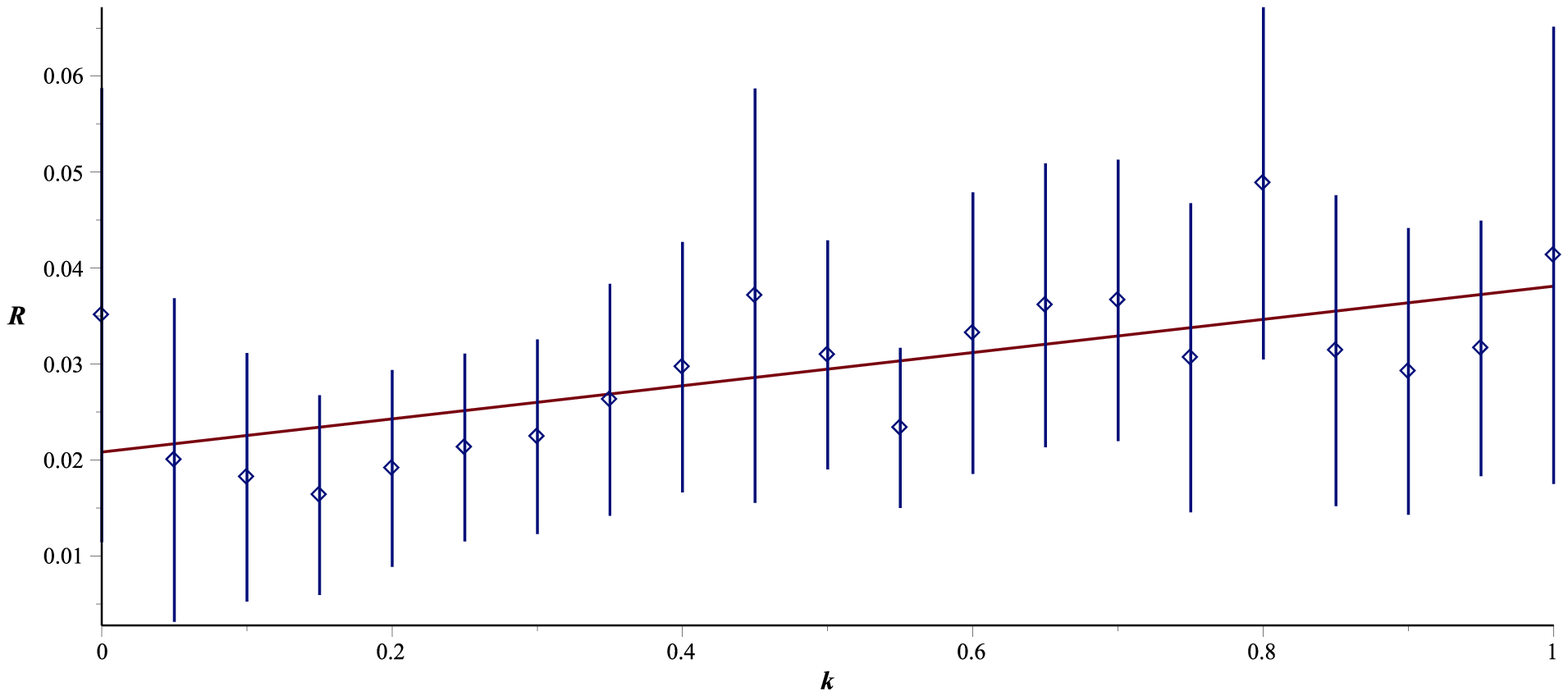} \\
(c)$\tau(k)$ for $M = 15$.  & (d)$R(k)$ for $M = 15$.\\[6pt]
 \includegraphics[width=65mm, height = 35mm]{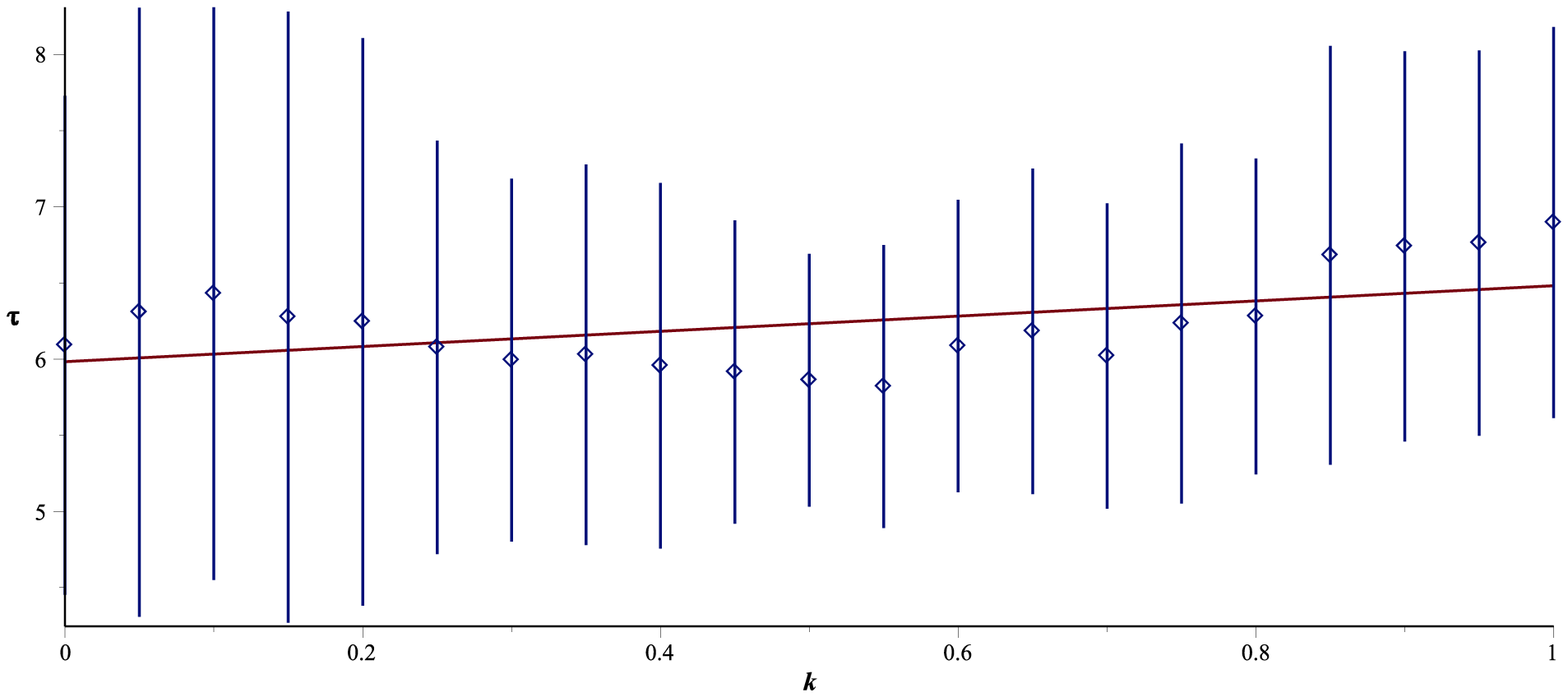} &   \includegraphics[width=65mm, height = 35mm]{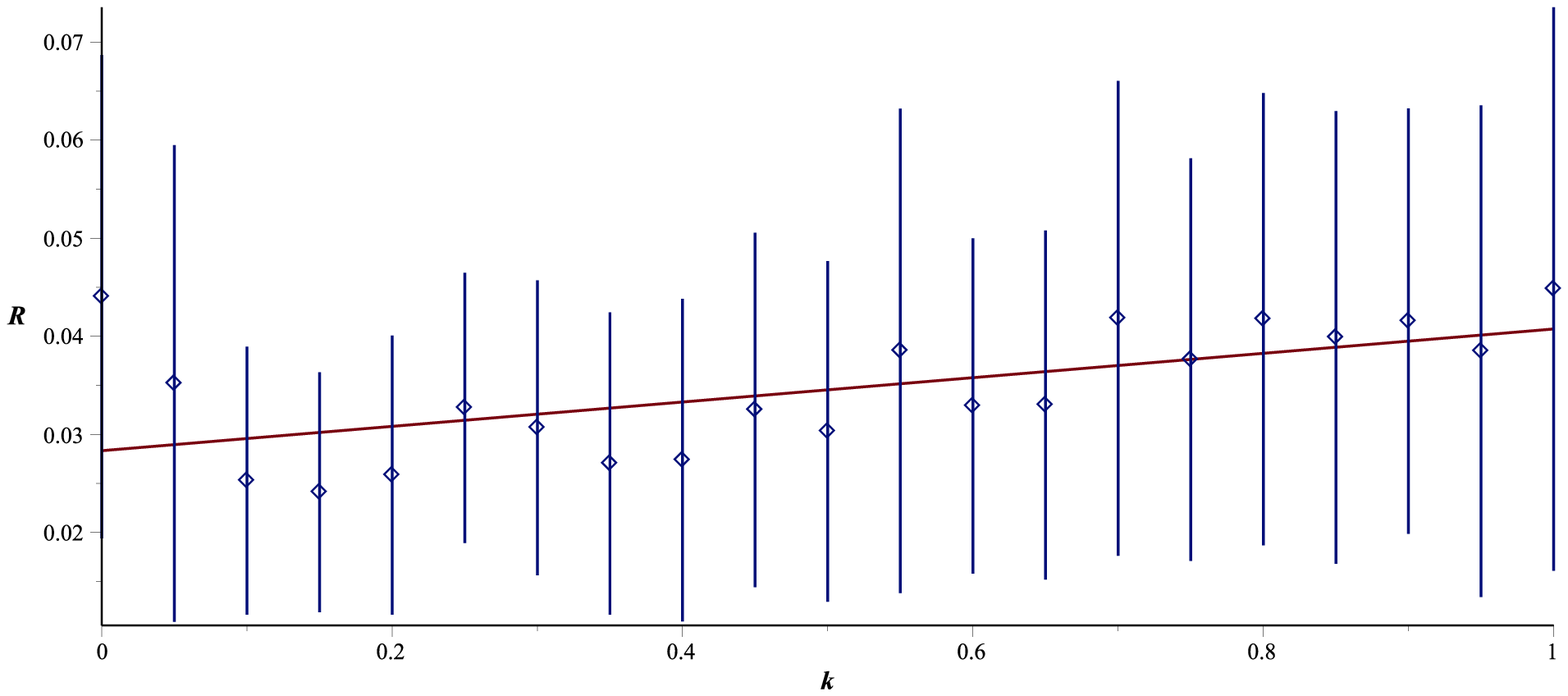} \\
(e) $\tau(k)$ for $M = 18$. & (f) $R(k)$ for $M = 18$.
\end{tabular}
\caption{Relaxation timescales and nonequilibrium residues as functions of the interaction coupling $k$ (varying from $0.00$ to $1.00$) obtained from the fittings of $\overline{H}(t)$. The values of $\tau$ and $R$ were calculated for each \textit{H}-function and then averaged over ten sets of randomly-chosen initial phases for each value of $k$. Error bars correspond to standard deviations over these ten values for each point.}
\end{figure}
\subsection{Relaxation Timescales, Residues and Interactions}

In the last two subsections we have used the Forward Trajectories Method to study the effect of the coupling constant on the evolution of distributions, trajectories and \textit{H}-functions showing how interactions might prevent nonequilibrium initial distributions from reaching a final equilibrium state. In order to properly quantify the effects indicated by those results we will now study how the \textit{H}-function evolves for different values of the coupling constant and the number of superposed modes of the wave function. In this subsection we study if and how interactions might be correlated to the relaxation timescales and possible nonequilibrium residues.

We have studied the evolution of 37 different \textit{H}-functions with increasing coupling constants ranging from $k = 0.0$ to $k=1.80$ during the time interval $t = [0, 12\pi]$. These \textit{H}-functions were calculated for ten different sets of random phases and then fitted to functions of the form $(\bar{H_0} - R)e^{-t/\tau} + R$ which in turn provided ten different values for the relaxation timescales $\tau$ and residues $R$. The values obtained from the best fits of the coarse-grained \textit{H}-functions were then averaged over and errors bars were obtained from the standard deviation of the ten different values for each $k$. The results of those simulations indicate the existence of two regimes concerning the variation of the coupling constant. The analysis of the evolution in those two regimes has led us to confirm the indications from the trajectories shown in Figure 4.

\begin{figure}[t]

\centering
\begin{tabular}{cc}
  \includegraphics[width=65mm, height = 35mm]{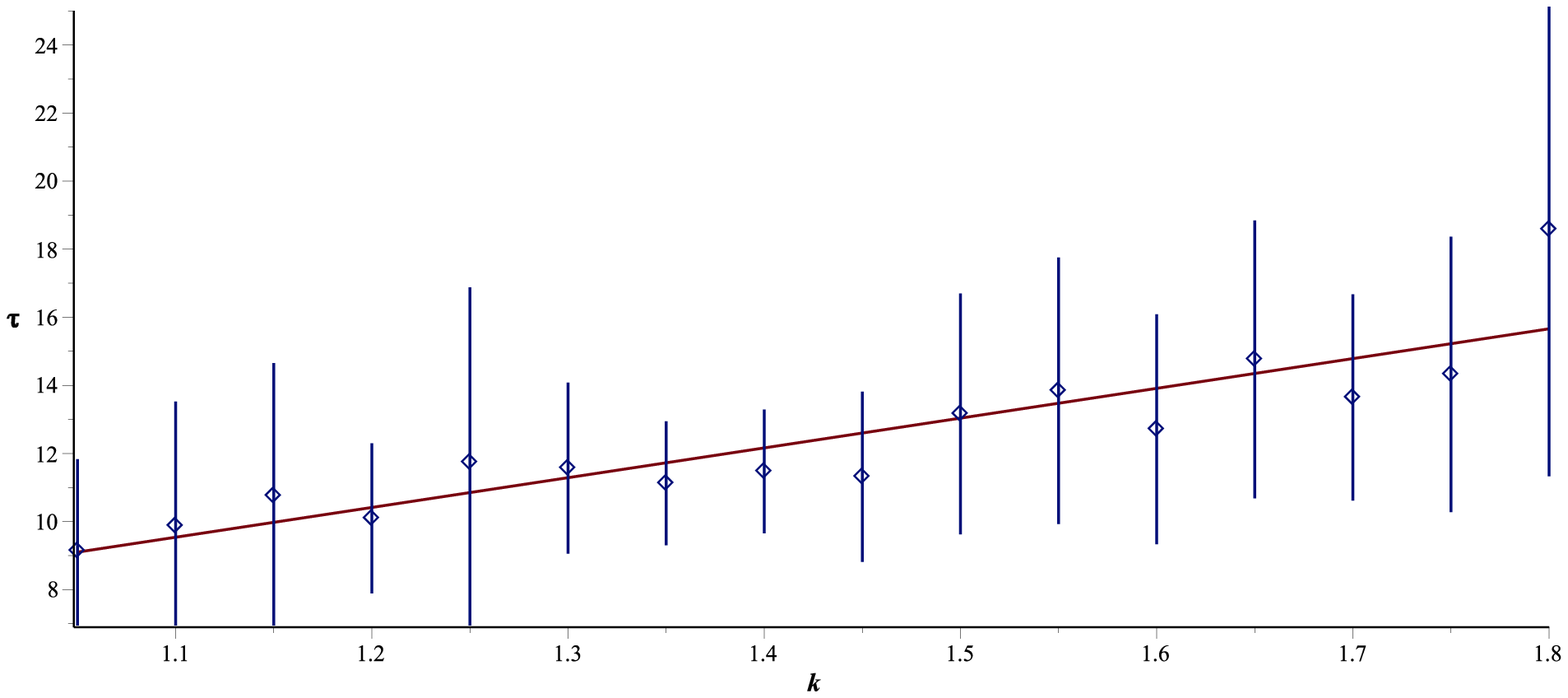} &   \includegraphics[width=65mm, height = 35mm]{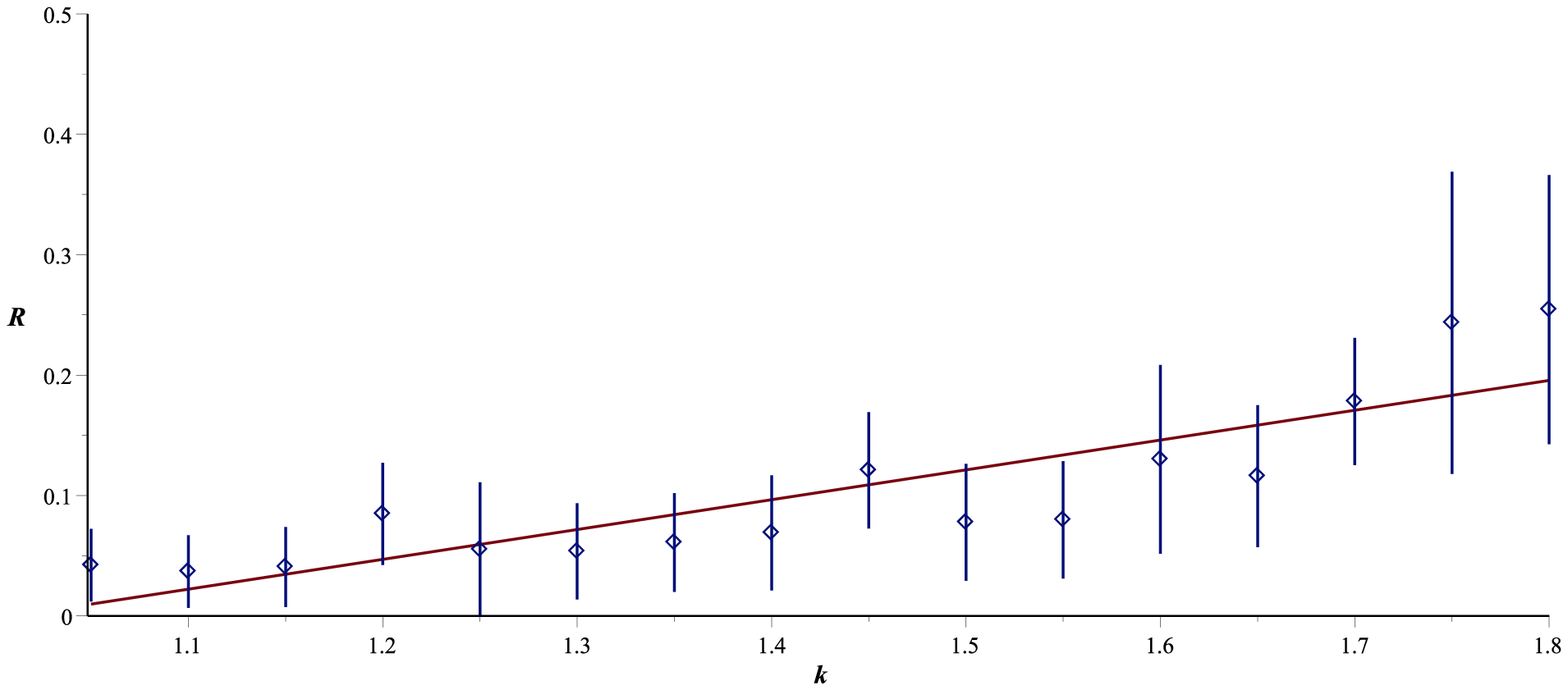} \\
(a)$\tau(k)$ for $M = 12$. & (b)$R(k)$ for $M = 12$.  \\[6pt]
 \includegraphics[width=65mm, height = 35mm]{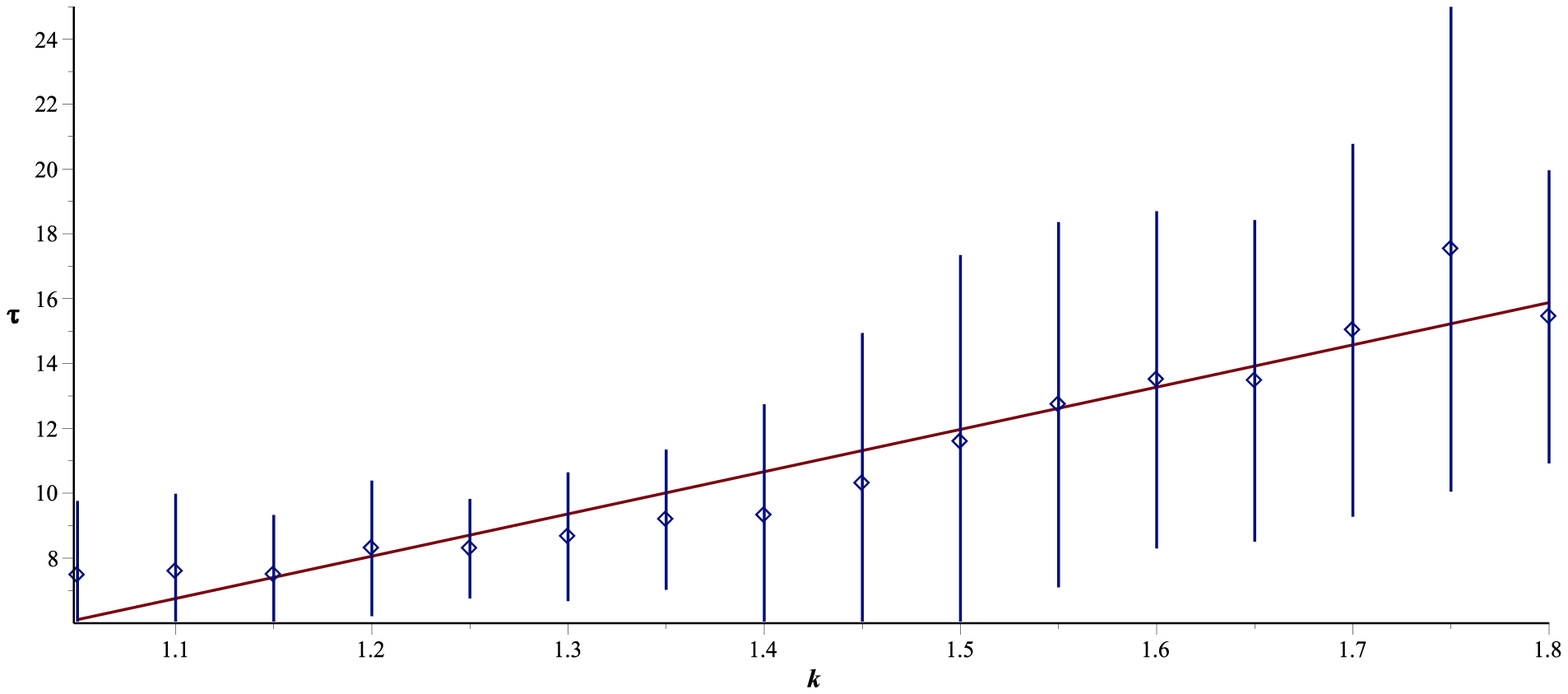} &   \includegraphics[width=65mm, height = 35mm]{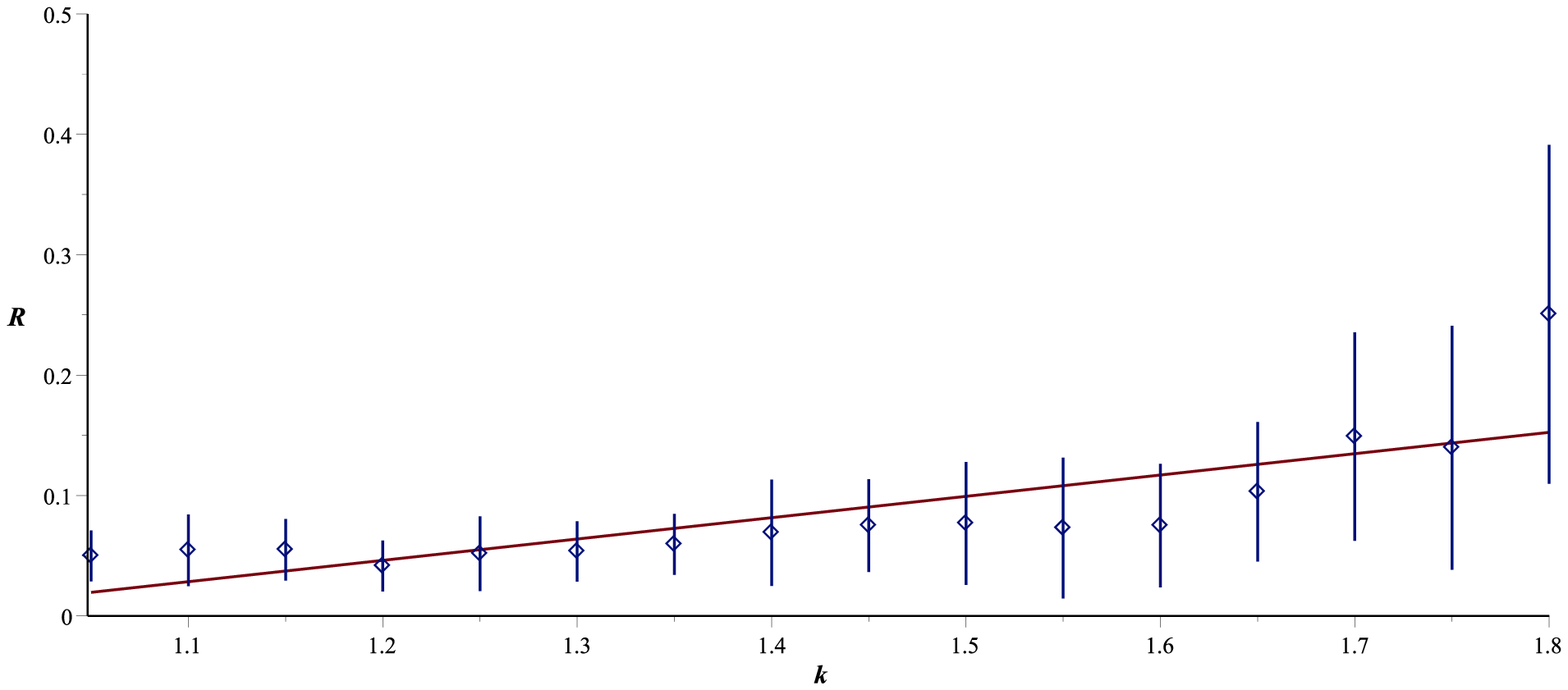} \\
(c)$\tau(k)$ for $M = 15$.  & (d)$R(k)$ for $M = 15$.\\[6pt]
 \includegraphics[width=65mm, height = 35mm]{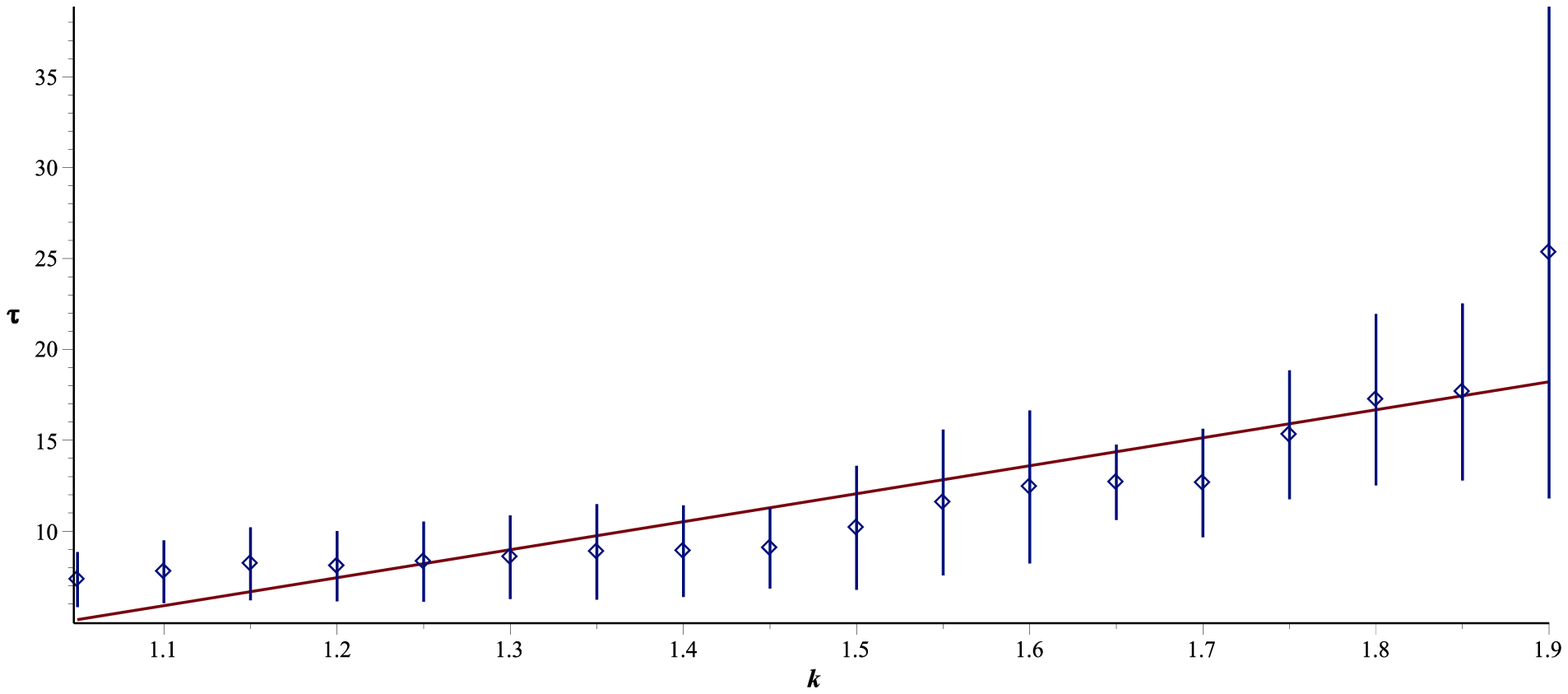} &   \includegraphics[width=65mm, height = 35mm]{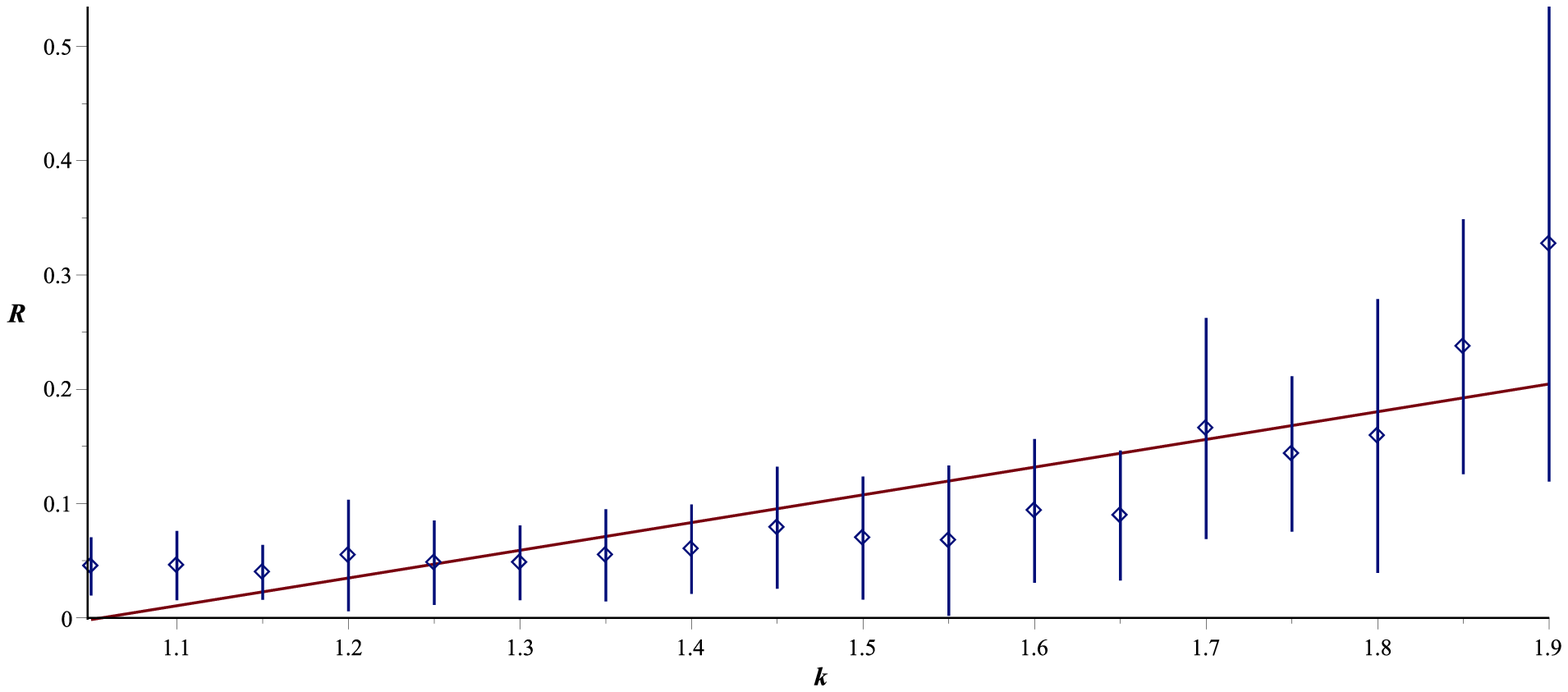} \\
(e) $\tau(k)$for $M = 18$. & (f) $R(k)$ for $M = 18$.
\end{tabular}
\caption{Relaxation timescales and nonequilibrium residues as functions of the interaction coupling $k$ (varying from $1.05$ to $1.80$) obtained from the fittings of $\overline{H}(t)$. The values of $\tau$ and $R$ were calculated for each \textit{H}-function and then averaged over the ten sets of random phases for each value of $k$. Error bars correspond to standard deviations over these ten values for each point.}
\end{figure}

This can be seen in Figures 6 and 7. In Figure 6 we observe a very slight increase in the values of both $\tau$ and $R$ and an overall oscillating evolution. This can be explained by the fact that the interaction does not dominate over the fundamental frequency ($k < \omega^2$) and its increase does not fundamentally affect the relaxation process. In Figure 7 the situation is quite different, the increase in the coupling constant is directly related to an increase in the relaxation timescale. The increase in the values of the nonequilibrium residue is also noticeable, although less accentuated.

These results indicate that, when the oscillating frequency $\omega$ dominates over the interaction, the effects of the coupling do not lead directly to an increase in the relaxation timescale or the nonequilibrium residue. However, when the interaction dominates it can significantly delay and perhaps even prevent complete quantum relaxation. This conclusion suggests that quantum nonequilibrium might survive situations where strong interactions dominate over other characteristics of the system. Such effects might occur in the interactions between scalar fields in the early universe or in the process of parametric resonance characteristic of inflationary reheating scenarios.

\section{Conclusions} \label{sec:conclusions}

In this paper we have studied the effects of interactions on quantum relaxation in a system of coupled one-dimensional harmonic oscillators. This study was done using a numerical Forward Trajectories Method that allowed us to analyze the system in a number of different ways. 

Our detailed study of the evolution of nonequilibrium distributions shows that relaxation  takes place efficiently for a significant number of superposed modes of the wave function and for a small value of the coupling constant $k$. Analysis of the trajectories shows that there is little or no confinement when the coupling is smaller then the squared fundamental frequency of the oscillators, indicating that in those cases relaxation is unlikelly to be affected by the coupling. When $k > \omega^2$ a higher percentage of trajectories stay confined to certain regions and complete relaxation is less likely,  although still possible after a long enough time and with a sufficient number of modes. 

The evolution of the coarse-grained \textit{H}-function for a long time period confirmed previous results which demonstrated that for a low number of modes ($M = 4$) quantum equilibrium  need not be reached exactly. For larger numbers of modes ($M=12$ and $M=20$) the system approximately obeys the Born rule after a sufficient time period. However, simulations with larger couplings show that the relaxation timescale grows significantly and nonequilibrium might still be large enough to produce detectable effects after a considerable time (for example, the value of $\overline{H}(t)$ represents more then $20\%$ of the initial value at $t=16\pi$ for the cases $M=12$ and $M=20$).

Finally, we analyzed the evolution of $\overline{H}(t)$ over a shorter time period $t = [0, 12\pi]$ for a variety of values of $k$, confirming the findings cited above. Fitting $\overline{H}(t)$ with a function of the form $(\bar{H_0} - R)e^{-t/\tau} + R$ we obtained the values of the timescale $\tau$ and of the nonequilibrium residues $R$ ranging from $k = 0.00$ to $k = 1.80$. The growth of the coupling does not significantly affect the relaxation parameters when $k < \omega^2$, indicating that when the interaction is not dominant it has no clear effect on the evolution of $\overline{H}(t)$. However, when $k > \omega^2$ the timescales increases  significantly with the increase in the coupling constant. The nonequilibrium residue also seems to be affected by the growth of $k$, but it is generally quite small for the numbers of modes considered $M =12, 15, 18$. In any case, the values of $R$ are generally higher when the interaction is stronger and this further confirms the indications of the rest of our analysis. 

Our numerical results provide a picture of how a simple interaction may delay or partially prevent quantum relaxation. Considering this as a toy model of more complicated systems that existed in the very early universe, where interactions might play an important role, the results suggest a way in which quantum nonequilibrium could survive the early stages of cosmological evolution. In standard inflationary scenarios, if some modes of the inflaton were still out of equilibrium after inflation, this would affect the `decay' of the inflaton field, which occurs via interactions with other fields and which creates particles in the early universe \cite{bib:Underwood2015}. If interactions play a role in preventing the relaxation to the Born rule, this could provide a feasible way through which nonequilibrium might survive the `reheating' phase.

Another relevant scenario would be to consider multi-field inflation \cite{bib:Wands2007, bib:Peter2016}. In such models it is still an open question how quantum nonequilibrium would evolve during inflation and after it. Nonetheless, in such models interactions play an important role and our results indicate that this would also be a situation where quantum relaxation could be delayed. 

Finally, in bouncing cosmological models (\cite{bib:Pinto-Neto2021} and references therein), it is possible that quantum vacuum fluctuations, originating in the far past of a contracting phase dominated by a pressureless matter field, generate almost scale invariant power spectra of cosmological perturbations. As the initial conditions are imposed in a distant past, it is plausible that relaxation had already occurred. However, if during a quantum bounce quantum gravity effects are significant, nonequilibrium could be generated during this phase \cite{bib:Valentini2021}. In the case of bouncing models with multiple fields where couplings are important in the evolution of perturbations, nonequilibrium may then propagate to the future, and the results presented here may be relevant to understand its evolution in the universe. All these hypotheses can be studied in depth using our numerical method if it is extended to more complex time-dependent wave functions. We hope to return to this in future work.

\textbf{DATA ACCESS:}\\
All the data used in our analysis were produced by original simulations developed by the
authors. The data used for the plots in our paper as well as the source codes of the simulations are accessible
through the link: (\url{https://drive.google.com/file/d/1XsYLu1VfaNH5yvmrLi29mKPhj8KHzbmA/view?usp=sharing}). The fitting of the data was made using mathematical software as described in the paper.

\textbf{AUTOR CONTRIBUTIONS:}\\
F.B.L.'s developed the code, produced the numerical results, worked on the numerical and theoretical analysis of the results and drafted the manuscript. N.P.N. worked on the theoretical analysis of the results and helped draft the manuscript. A. V. also worked on the theoretical analysis of the results and on the draft of the manuscript.

\textbf{FUNDING:}\\F.B.L.'s work was funded by CNPq (PCI-DC grant 300769/2022-9 associated with the PCI/MCTI/CBPF program). N.P.N. acknowledges the support of CNPq of Brazil under grant PQ-IB 310121/2021-3.

\textbf{AKNOWLEDGEMENTS:}\\F.B.L. would like to acknowledge the fundamental contribution of Samuel Colin in the development of the numerical method used in our work.


\bibliography{mainexemplos} 

\end{document}